\documentclass[nofootinbib,aps]{revtex4}

\usepackage{amstext,amsmath,amssymb,amsfonts,bbm}
\usepackage{amsthm}
\usepackage[dvips]{graphicx}
\usepackage{latexsym}
\usepackage{psfrag}

\def \v{\vspace}
\newcommand{\Ref}[1]{(\ref{#1})}
\def \subs{\subsection}

\newcommand{\N}{\mathbb{N}}
\newcommand{\Z}{\mathbb{Z}}

\newcommand{\R}{\mathbb{R}}
\newcommand{\C}{\mathbb{C}}

\DeclareMathOperator{\tr}{tr}

\theoremstyle{definition}

\newtheorem{lemma}{Lemma}

\newtheorem{defi}{Definition}
\newtheorem{expl}{Example}

\def\be{\begin{equation}}
\def\ee{\end{equation}}
\def\bes{\begin{eqnarray}}
\def\ees{\end{eqnarray}}

\def\arr{\rightarrow}

\def\om{\omega}

\def\la{\langle}
\def\ra{\rangle}
\def\f{\frac}
\def\tl{\tilde}
\def\wt{\widetilde}
\def\what{\widehat}

\def \vphi{\varphi}

\newcommand{\lalg}[1]{\mathfrak{#1}}  

\newcommand{\su}{\lalg{su}}

\renewcommand{\v}{\overrightarrow}
\def\dag{^\dagger}
\newcommand{\SU}{\mathrm{SU}}

\def\pp{\partial}
\def\hh{{\cal H}}
\def\v{\vec}

\def\Id{{\mathbbm 1}}

\def\6{\langle}
\def\9{\rangle}
\def\tr{{\rm tr}\,}

\def\1{\mbox{1\hskip-.25em l}}

\newcommand{\ket}[1]{|#1\rangle}

\def\ii{{\cal I}}
\def\jj{{\cal J}}
\def\ss{{\cal S}}
\def\d1{{}^1 d}

\def\aa{{\cal A}}
\def\vv{{\cal V}}
\def\Ll{{\cal L}}

\def\Int{{\rm Int}}

\def\Ee{{\cal E}}
\def\vio{{\vec{\ii}}}
\def\vjo{{\vec{\jj}}}

\def\mm{{\cal M}}
\def\ext{{\rm ext}}
\def\inte{{\rm int}}

\begin{document}
\title{Reconstructing Quantum Geometry from Quantum Information: \\
Area Renormalisation, Coarse-Graining and Entanglement on Spin
Networks}

\author{{\bf Etera R. Livine}\footnote{elivine@perimeterinstitute.ca}}
\affiliation{Perimeter Institute, 31 Caroline St N., Waterloo, ON, Canada N2L 2Y5}
\affiliation{Laboratoire de Physique, ENS Lyon, CNRS UMR 5672, 46 All\'ee d'Italie, 69364 Lyon Cedex 07, France}
\author{{\bf Daniel R. Terno}\footnote{dterno@perimeterinstitute.ca}}
\affiliation{Perimeter Institute, 31 Caroline St N, Waterloo, ON, Canada N2L 2Y5}

\begin{abstract}
\begin{center} {\small ABSTRACT} \end{center}

After a brief review of spin networks and their interpretation as
wave functions for the (space) geometry, we discuss the
renormalisation of the area operator in loop quantum gravity. In
such a background independent framework, we propose  to probe the
structure of a surface through the analysis of the coarse-graining
and renormalisation flow(s) of its area. We further introduce a
procedure to coarse-grain spin network states and we
quantitatively study the decrease in the number of degrees of
freedom during this process. Finally, we use these coarse-graining
tools to define the correlation and entanglement between parts of
a spin network and discuss their potential interpretation as a
natural measure of distance in such a state of quantum geometry.

\end{abstract}

\maketitle

\tableofcontents


\section{Introduction}

Loop Quantum Gravity (LQG) proposes a background independent
framework for a theory of quantum general relativity \cite{lqg}.
It realizes a canonical quantization of general relativity in 3+1
space-time dimensions and defines the Hilbert space of quantum
states of 3d space geometry and their dynamics (through a
Hamiltonian constraint). Background independence means that there
is no assumed background metric at all and that the quantum state
of geometry describes the whole metric of space(-time) and not
simply the perturbations of the metric field around a fixed
background metric. The whole geometry of the space(-time) needs to
be reconstructed from the quantum state: all geometric notions,
such as the distance, need to be constructed from scratch since
you can not rely on a background geometry to define, for instance,
a reference notion of distance that you could use to describe the
metric perturbations.

The states of 3d space geometry are (superpositions) of spin
networks. These are equivalence classes of labelled graphs under
spatial diffeomorphism. The labels come from the representation
theory of the gauge group $\SU(2)$, representation labels on the
edges and invariant tensors (or singlet states) at the vertices.
These spin networks thus contained both algebraic (or
combinatorial) data and topological data such as winding numbers
of the edges of the graph. In the present paper, we will neglect
this topological information and work with {\it abstract spin
networks} which are purely algebraic objects. This erases all
traces of the topology of the original 3d space manifold which we
quantize. This is an open issue in LQG: whether the quantum states
of geometry describe both the metric and topology of space or
whether they should only describe the metric state in some fixed
background (space) topology. The standard LQG framework fixes the
background topology, uses embedded spin networks and thus forbids
topology changes, whereas the more recent spin foam framework (see
for example \cite{spinfoam}), which proposes a path integral
formalism for LQG, uses abstract spin networks and allows
transitions between different spatial topologies. More generally,
it seems possible to extract some topological information from the
algebraic data encoded in the spin network (see for example
\cite{flo}). However, on the other hand, the topological degrees
of freedom allowed in spin networks leads to interesting
proposals, such as a possible inclusion of the standard model for
particles in LQG \cite{sundance}.

It is essential to better understand the quantum geometry defined
by spin networks and how a classical metric can emerge. A first
step would be to develop the concept of {\it distance} on a spin
network. Since we are in a completely background independent
context and that the spin network ought to define the (quantum)
state of geometry, the algebraic and combinatorial structures of
the spin networks must induce a (natural) notion of distance on
the ``quantum manifold" defined by the spin network. Moreover,
since there is no background metric defining a reference frame,
there is no concept of the ``position" of a part of the spin
network and we have to deal with {\it relations} between parts of
the spin network. It is therefore natural to try to reconstruct a
notion of distance between two parts of the spin network as a
function of the correlations between these parts, as suggested in
\cite{flo}.

Developing a notion of distance, close and far, is a necessary
tool to defining a proper coarse-graining procedure of the spin
network, which would not rely on an assumed embedding of the spin
network in a fixed background manifold but solely on the spin
network state itself. The goal is not to assume but to derive the
embedding of the spin network into a classical manifold, which
would define its semi-classical limit. Such coarse-graining would
thus allow to study the emergence of a classical metric in LQG and
the semi-classical dynamics of the theory.

\medskip

In the next section, we introduce the mathematics of the spin
networks and we describe the quantum states of geometry defined by
LQG. We insist that the spin networks are wave functions of the
geometry/metric and we remind their interpretation as probability
amplitudes.

In section \ref{AreaRenorm}, we remind the definition of surfaces
in LQG and describe the (quantum) surface states. Surfaces can be
thought as made of elementary quantum surfaces patched up
together. We define a coarse-graining of these quantum surfaces in
the spirit of the area renormalisation introduced in \cite{ourbh}.
We explain how renormalisation flow reflects the structure of
surface i.e how the elementary surfaces are patched together to
form the big surface.

In section \ref{Coarse}, we introduce a notion of coarse-grained
state of a bounded region of the spin network and we discuss the
properties of the coarse-grained state. More details on the
relation between the coarse-graining and the usual operation of
partial tracing is presented in appendix. In section
\ref{Entanglement}, we use that tool in order to define the
entanglement and correlation between two parts of the spin
network. Finally in section \ref{TheIdea}, we argue that these
correlations define a natural notion of distance between parts of
the spin network. This distance would be determined solely from
the algebraic structure of the spin network and would not rely on
any background structure. Such a relationship between geometrical
quantities (such as the distance or the metric) and informational
concepts (such correlations) supports the growing link between
(loop) quantum gravity and quantum information
\cite{flo,ourbh,fotini,dt}.

\section{Spin Networks et cetera}

\subs{Quantum states of geometry and intertwiner space}

Loop quantum gravity is a canonical formulation of general
relativity, describing the (quantum) evolution of the 3d metric
$h_{ij}(x)$ on a spatial slice $\Sigma$ of space-time. The
Ashtekar-Barbero variables describing the 3d geometry are a
$\SU(2)$-connection $A_i^a(x)$ and a triad $E_i^a(x)$
($\su(2)$-valued 1-form) which is its canonical momentum. The
connection $A$ describes the parallel transport in the 3d space,
while $E$ defines the 3d metric via $h_{ij}=E_i^aE_j^a$. In these
variables, general relativity becomes a $\SU(2)$ gauge theory,
with Gauss constraints insuring the $\SU(2)$ gauge invariance and
other constraints implementing the invariance under (space-time)
diffeomorphisms.

Loop quantum gravity chooses the $A$-polarization. We define wave
functions $\psi(A)$ and $E$ acts as a derivation operator. We
select {\it cylindrical functions}, which depend on the field $A$
through a finite number of parameters. A cylindrical function is
defined with respect to an oriented (closed) graph $\Gamma$
(embedded in $\Sigma$). Let us construct the holonomies
$U_e[A]\in\SU(2)$ of the connection $A$ along the edges $e$ of
$\Gamma$. A cylindrical function on $\Gamma$ is defined as a
function of these holonomies:
$$
\psi_\Gamma(A)=\psi(U_e[A]), \quad e\in\Gamma.
$$
Moreover, we require that these functions are $\SU(2)$ gauge
invariant i.e invariant under gauge transformations of the
connection $A\arr g^{-1}Ag +g^{-1}{\rm d}g$ for $g(x)\in\SU(2)$.
This translates into the requirement of $\SU(2)$ invariance at
every vertex $v$ of the graph $\Gamma$:
\be
\psi(U_e[A])=\psi(g_{s(e)}^{-1}U_e[A]g_{t(e)}), \quad
\forall g_v\in\SU(2),
\ee
where $s(e)$ and $t(e)$ denote the source and target vertices of
the edge $e$. Therefore, for a fixed graph $\Gamma$, we are
looking at the space of functions over $\SU(2)^E/\SU(2)^V$ where
$E$ and $V$  are the number of edges and vertices of $\Gamma$
respectively. This space is provided with the Haar measure on
$\SU(2)^E$, which defines the kinematical\footnote{The physical
scalar product should take into account the Hamiltonian
constraint, which generates diffeomorphisms along the time
direction and dictates the dynamics of the theory.} scalar product
of LQG. This measure defines a Hilbert space
$H_\Gamma=L^2(\SU(2)^E/\SU(2)^V)$ of wave functions associated to
the graph $\Gamma$. The invariance of the theory under spatial
diffeomorphisms is taken into account by considering equivalence
classes of graphs on $\Sigma$ under diffeomorphisms. The
(kinematical) Hilbert space of LQG, $H_{\rm kin}$, is then
constructed as the projective limit of the $L^2$ spaces associated
to each (equivalence class of) graph, which implements a sum over
all possible graphs. For more details, the interested reader can
refer to \cite{lqg}.

\medskip

A basis of $H_\Gamma=L^2(\SU(2)^E/\SU(2)^V)$ is provided by the
so-called {\it spin network} functionals. These are constructed as
follows.  To each edge $e\in\Gamma$, we associate a $\SU(2)$
representation labelled by a half-integer $j_e\in\N/2$ called
{\it spin}. The representation (Hilbert) space is denoted $V^{j_e}$
and has a dimension $d_{j_e}=2j_e+1$. To each vertex $v$, we
attach an intertwiner $\ii_v$, which is $\SU(2)$-invariant map
between the representation spaces $V^{j_e}$ associated to all the
edges $e$ meeting at the vertex $v$:
$$
\ii_v:\,\bigotimes_{e\,{\rm ingoing}} V_{j_e}\arr \bigotimes_{e
\,{\rm outgoing}} V_{j_e}.
$$
Since the conjugate representation $\overline{V_{j}}$ is
isomorphic to the original $V_{j}$, we can alternatively consider
$\ii_v$ as a map from $\bigotimes_{e\ni v} V^{j_e}$ to $\C\cong
V^0$. Then one can also call the intertwiner $\ii_v$ an
{\it invariant tensor} or a {\it singlet state}.
Once the $j_e$'s are fixed, the intertwiners at the vertex $v$
actually form a Hilbert space, which we will call
$\Int_v\equiv\Int(\bigotimes_{e\ni v} V_{j_e}\arr\C)$. Moreover,
considering the  decomposition into irreducible representations of
the tensor product $\bigotimes_{e} V_{j_e}$,
$$
\bigotimes_{e\ni v} V_{j_e}=\bigoplus_j \alpha_j^{\{j_e\}} V^j,
$$
where the $\alpha_j$ are degeneracy coefficients depending on the
$j_e$'s, we call $\hh^j_v$ the subspace corresponding to the spin
$j$ component of the tensor decomposition. Then $\Int_v$ is
actually the space of singlets $\hh^0_v$ corresponding to the spin
$j=0$ component.

A spin network state  $|\Gamma,\{j_e\},\{\ii_v\}\ra$ is defined as
the assignment of representation labels $j_e$ to each edge and the
choice of a vector $|\{\ii_v\}\ra \in \bigotimes_v\Int_v$ for the
vertices. To shorten the expressions, we may omit the indices and
use vectorial notations, $|\Gamma,\vec{\jmath},\vio\ra$. The spin
network state defines a wave function on the space of discrete
connections $\SU(2)^E/\SU(2)^V$,
\be
\label{spinnetdef}
\phi_{\vec{\jmath},\vio}[g_e]
= \la g_e|\ii_v \ra
\equiv \tr \bigotimes_e D^{j_e}(g_e) \otimes \bigotimes_v \ii_v,
\ee
where we contract the intertwiners $\ii_v$ with the (Wigner)
representation matrices of the group elements $g_e$ in the chosen
representations $j_e$. Using the orthogonality of the
representation matrices for the Haar measure,
$$
\int_{\SU(2)}dg\,
\overline{D^{j}_{ab}(g)}D^{k}_{cd}(g)=
\f{\delta_{jk}}{d_j}\delta_{ac}\delta_{bd},
$$
we can directly compute the scalar product between two spin
network states:
\be
\label{matnorm}
\la\Gamma,\vec{k},\vjo|\Gamma,\vec{\jmath},\vio\ra
\,=\,
\prod_e \f{\delta_{j_ek_e}}{d_{j_e}}\,
\prod_v \la\jj_v|\ii_v\ra.
\ee
Therefore, upon choosing a basis of intertwiners for every
assignment of representations $\{j_e\}$, the spin networks provide
a basis of the space $H_\Gamma=L^2(\SU(2)^E/\SU(2)^V)$. Summing
over all graphs $\Gamma$, we then define the spin network basis of
LQG.

\medskip

Considering that the triad $E$ allows to reconstruct the metric
$g\sim E^2$ and implementing $E$ as a derivation operator on the
wave functions, one can compute the action of geometrical
operators such as the area and the volume on the spin network
basis.

For instance, considering a graph $\Gamma$ and a spin network
state $|\Gamma,\vec{\jmath},\vio\ra$, the area of an elementary
surface $\ss_e$ transversal to a given edge $e$ (and not
intersecting any other edges of the graph) will have a quantized
spectrum:
\be
\what{\aa}_{\ss_e}\,|\Gamma,\vec{\jmath},\vio\ra
\,=\,
l_P^2 S(j_e)|\Gamma,\vec{\jmath},\vio\ra,
\ee
where $S(j)=\sqrt{j(j+1)}$ or $\tl{S}(j)=j$ upon a choice of
ordering in the quantization. Note that $S(j)=\sqrt{j(j+1)}$ is
the value of the Casimir operator $\sqrt{\v{J}^2}$ in the
representation $V^j$, while
$\tl{S}(j)=j=\sqrt{\v{J}^2+\f14}-\f12$. One can also investigate
the action of the volume operator and it would be seen to act only
at the vertices of the graph and depend on the intertwiners
$\ii_v$. To sum up the situation, the representation labels can be
considered as the quantum numbers for the area while the
intertwiners would be the quantum numbers for the (3d) volume.
This provides the spin network states with an interpretation as
{\it discrete geometries}.

\medskip

We conclude this review section with a few points on the
representation theory of $\SU(2)$. For a given representation
$V^j$, the trace of the representation matrices $D(g)$ defines the
character $\chi_j:\SU(2)\arr \C$,
$$
\chi_j(g)\,=\,\tr_j\left(D^j(g)\right)
=\f{\sin(2j+1)\theta}{\sin\theta},
$$
where $g=\exp(i\theta\hat{u}.\v{\sigma})=\cos\theta\,\Id+i
\sin\theta\,\hat{u}.\vec{\sigma}$, $\theta\in[0,2\pi]$ is
half the angle of the rotation, $\hat{u}\in\ss^2$ is a unit vector
on the 2-sphere indicating the axis of the rotation and
$\v{\sigma}$ are the standard Pauli matrices. We will write in the
following $g=(\theta,\hat{u})$. Note that the $\chi_j$'s are
invariant under conjugation $g\arr h^{-1}gh$.

In these variables, the normalized Haar measure on $\SU(2)$
reads\footnote{The Haar measure on $\SU(2)$ is actually the usual
Lebesgue measure on the 3-sphere $\ss^3$. Indeed with the
parametrization $g=\pi_0
\Id + i \pi_i \sigma^i$, we have:
$$
dg\,=\,d^4\pi_\mu\,\delta(\pi_\mu\pi^\mu-1).
$$}:
$$
\int_{\SU(2)}dg\,f(g)=\,
\f1{2\pi^2}\int_{0}^{\pi}\sin^2\theta\,d\theta\,
\int_{\ss^2}d^2\hat{u}\,
f(\theta,\hat{u}).
$$
Note that $g(\theta,\hat{u})=g(-\theta,-\hat{u})$ and that the
angle $\theta$ is defined modulo $2\pi$.

Finally, we remark that the following operator is actually the
identity on the intertwiner space $\Int(j_1\otimes..\otimes j_n)$:
\be
\int_{\SU(2)} dg\, \bigotimes_{i=1}^n D^{j_i}(g)
\,=\,\Id.
\label{identity}
\ee
This defines the maximally mixed state on the singlet space
$\Int(j_1\otimes..\otimes j_n)$ up to a normalization factor.

\subs{Spin networks as probability amplitudes}

Even though spin network states diagonalize geometrical operators
such as areas and volumes, one must not forget their definition as
wave functions: they define probability distributions on the space
on (discrete) connections. More precisely, a normalized
cylindrical function $\psi_\Gamma$ defines a probability measure
on $\SU(2)^E/\SU(2)^V$:
\be
p(\{g_e,e\in\Gamma\})\,\prod_e dg_e
\,\equiv\,
|\psi_\Gamma(g_e)|^2 \,\prod_e dg_e, \qquad
\int_{\SU(2)^E}  \prod_e dg_e \, |\psi_\Gamma(g_e)|^2=1.
\ee
This way, spin networks also contain information about the
parallel transport on the 3d manifold.

We give two simple examples to illustrate the use of the spin
network as a probability amplitude.

\begin{expl}
Consider the simplest spin network graph based on a single closed
loop (with a single vertex). The spin network wave function is
labelled by a single representation label $j$ (and no intertwiner
is needed) and is defined as:
\be
W_j(g)=D^j_{mn}(g)\delta_{mn}=\chi_j(g).
\ee
$W_j(g)$ is invariant under conjugation and the only gauge
invariant data is the angle $\theta$ of the rotation defined by
$g$. Actually $W_j$ is the usual {\it Wilson loop}.

The character is already normalized $\int dg\,|\chi_j(g)|^2=1$ so
that $W_j$ define the following probability distribution:
\be
p(\theta)\,d\theta\,=\,
\sin^2\theta\, |\chi_j(\theta)|^2
=\sin^2 (2j+1)\theta,
\ee
where the $\sin^2\theta$ factor comes from the Haar measure. This
probability distribution is $\pi$-periodic. Its maxima -the most
probable parallel transport along the loop- are given by
$\theta=(\pi/2+n\pi)/d_j$, $n\in\Z$. On the other hand,
$p(\theta)$ vanishes for $\theta=n\pi/d_j$ and in particular for
the flat connection $\theta=0$.

The states $W_j$ diagonalize the area of an elementary surface
transversal to the loop. The area spectrum is $S(j)\,l_P^2$ as
states above. These states are in a sense completely delocalized
in $\theta$. On the other hand, one can write states
$\delta_\theta(g)$ localizing the rotation angle of the holonomy
$g$ defined as the following distributions:
$$
\int dg\, \delta_\theta(g) f(g)
=\f{1}{4\pi}\int_{\ss^2}d^2\hat{u} f((\theta,\hat{u})).
$$
We can decompose these states in the $W_j$ basis:
\be
\delta_\theta(g)\,=\,
\sum_{j\in\N/2} \chi_j(\theta)\chi_j(g).
\ee
They have an infinite expansion in the spin network basis and are
not normalisable. One could write normalized states defined by
Gaussian packets centered around fixed values of $\theta$ to
smooth out the state or by adding some regularization factors like
$\exp(-\kappa d_j)$, and study their $j$-average and their
properties, but this is not the purpose of the present paper.

\end{expl}

\begin{expl}
The $\Theta$ diagram has a more interesting structure of maxima.
Its has two vertices linked by three edges. Its spin network wave
functional is determined by three representation labels
$j_1,j_2,j_3$ and is defined as:
\be
\Theta_{j_1,j_2,j_3}(g_1,g_2,g_3)
\,\equiv\,
\sqrt{d_{j_1}d_{j_2}d_{j_3}}\,
C^{j_1j_2j_3}_{m_1m_2m_3}C^{j_1j_2j_3}_{n_1n_2n_3}\,
\prod_{i=1}^3 D^{j_i}_{m_i n_i}(g_i),
\ee
where the $C$'s are the (normalized) Clebsh-Gordan coefficients of
the recoupling theory of $\SU(2)$ representations. We can re-write
this expression in terms of the characters\footnotemark:
\be
\Theta_{j_1,j_2,j_3}(g_1,g_2,g_3)
\,=\,
\sqrt{d_{j_1}d_{j_2}d_{j_3}}\,
\int_{\SU(2)} dh\,
\prod_{i=1}^3 \chi_{j_i}(g_ih).
\ee
\footnotetext{We use the following integral definition of the Clebsh-Gordan coefficients:
$$
\int dh\,\prod_{i=1}^3 D^{j_i}_{m_in_i}(h)\,=\,
C^{j_1j_2j_3}_{m_1m_2m_3}C^{j_1j_2j_3}_{n_1n_2n_3}.
$$
This follows directly from eqn.\Ref{identity} for a 3-valent
vertex, since the space of 3-valent intertwiners is actually
one-dimensional.} It is straightforward to check that the norm of
$\Theta$ is $||\Theta_{j_i}||^2=\int dh \,\prod_{i}
\chi_{j_i}(h)$, which is either 1 is the $j_i$'s satisfy the
triangular inequality (so that $C^{j_1j_2j_3}\ne 0$) or 0
otherwise.

$\Theta$ is invariant under diagonal (left and right) $\SU(2)$
multiplication:
$$
\Theta(g_1,g_2,g_3)=\Theta(hg_1,hg_2,hg_3)=\Theta(g_1h,g_2h,g_3h),\,
\forall h.
$$
Using this $\SU(2)$ gauge invariance, we can define the loop
variables $\wt{g}_{1,2}=g_{1,2}g_3^{-1}$ and we have:
$$
\Theta(g_1,g_2,g_3)=\Theta(\wt{g}_1,\wt{g}_2,\Id)
=\sqrt{d_{j_1}d_{j_2}d_{j_3}}\,
\int_{\SU(2)} dh\,\chi_{j_1}(\wt{g}_1h)\chi_{j_2}(\wt{g}_2h)\chi_{j_3}(h).
$$
Writing $\wt{g}_1=(\theta_1,\hat{u}_1)$ and
$\wt{g}_2=(\theta_2,\hat{u}_2)$, it is easy\footnotemark to notice
that the previous expression only depends on $\theta_1,\theta_2$
and the angle $\varphi$ between $\hat{u}_1$ and $\hat{u}_2$.
\footnotetext{
Since $\chi_{j_3}(h)$ is invariant under conjugation $h\arr
k^{-1}hk$, we can do a change of variables on $h$ to rotate
$\wt{g}_1=(\theta_1,\hat{u}_1)$ to
$G_1=k\wt{g}_1k^{-1}=(\theta_1,\hat{u}_0)$ where
$\hat{u}_0=(0,0,1)$ marks the north pole on the 2-sphere. $k$ is
actually defined up to a rotation $r$ around the $z$-axis, $k\arr
rk$. We can use this final ambiguity to rotate
$\wt{g}_2=(\theta_2,\hat{u}_2)$ to
$G_2=rk\wt{g}_2(rk)^{-1}=(\theta_2,\hat{v}_2)$ with
$(\hat{v}_2)_y=0$, $\hat{v}_2=(\sin\varphi,0,\cos\varphi)$.}
 The angles $(\theta_1,\theta_2,\varphi)$ represent all the
gauge-invariant data for the discrete connections $(g_1,g_2,g_3)$
defined on the $\Theta$-graph. Taking into account the Haar
measure and the measure on $\ss^2$, the probability distribution
on the gauge-invariant space is:
\be
p(\theta_1,\theta_2,\varphi)
\, d\theta_1d\theta_2d\varphi
\,=\,
\f{1}{8\pi^2}
|\Theta(\wt{g}_1,\wt{g}_2,\Id)|^2
\sin^2\theta_1\sin^2\theta_2\sin\varphi\,
d\theta_1d\theta_2d\varphi.
\ee
The resulting probability distributions in the case of
$j_1=j_2=j_3=2,3,4,6$ are shown on
figs.~\ref{figtheta}-\ref{figtheta3}.



\begin{figure}[t]
\begin{center}
\includegraphics[width=8cm]{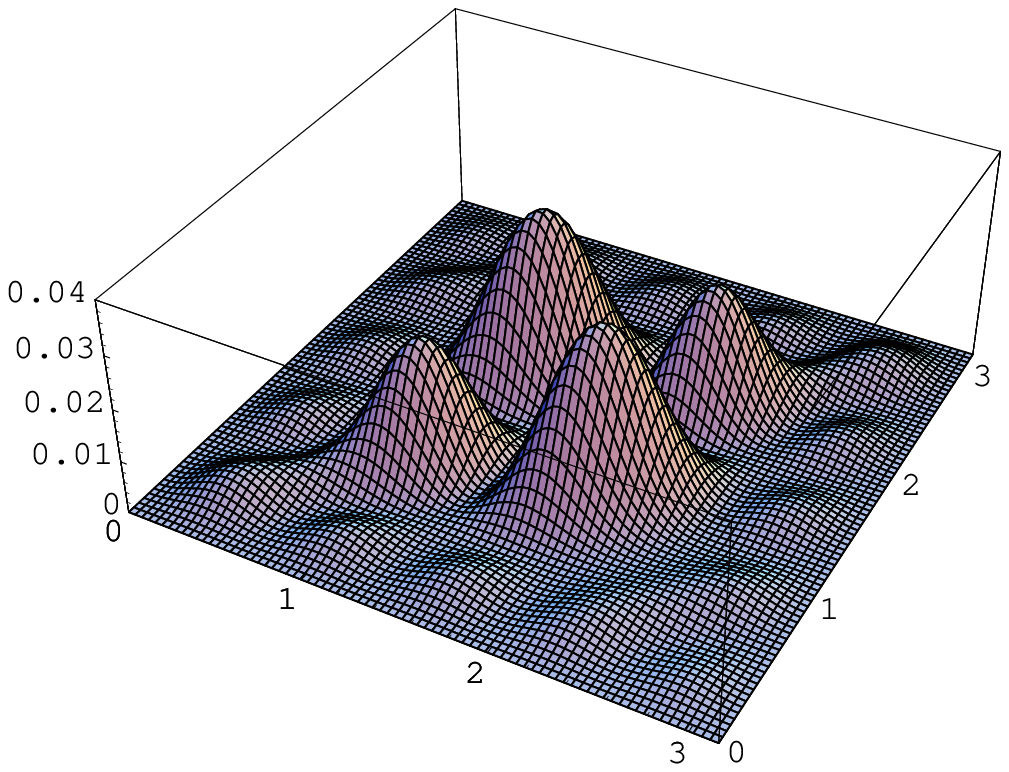}
\includegraphics[width=8cm]{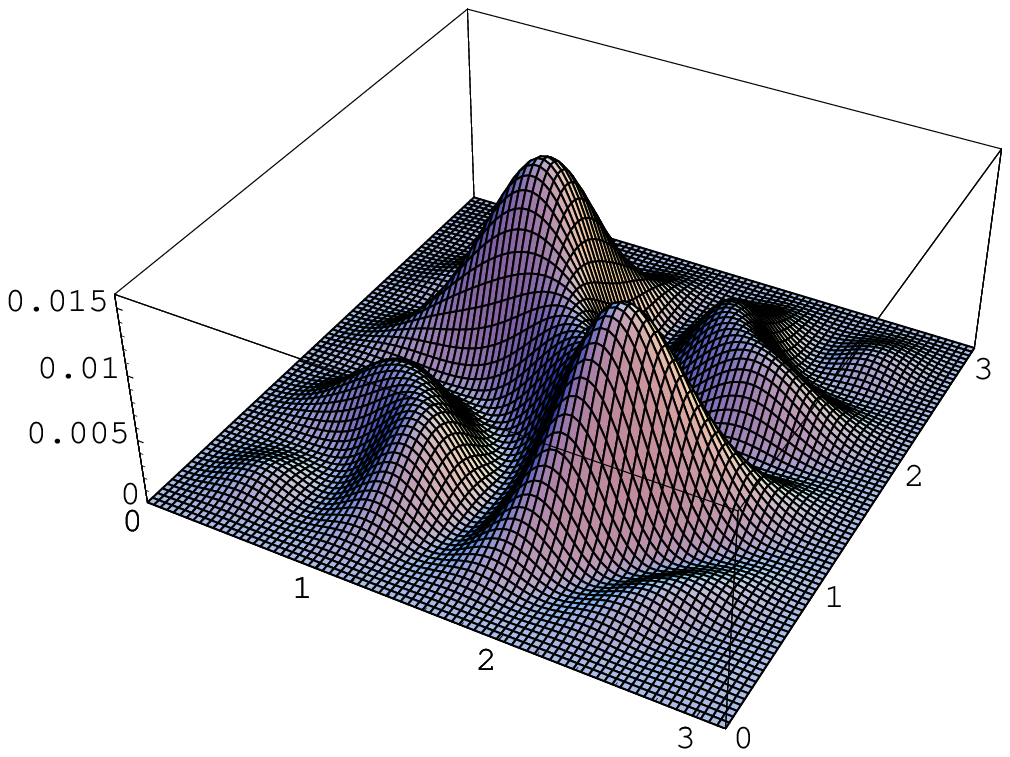}
\end{center}
\caption{\label{figtheta}
\small{Probability distribution
$p(\theta_1,\theta_2,\varphi=\pi/2)$ and
$p(\theta_1,\theta_2,\varphi=\pi/21)$ for $j_1=j_2=j_3=2$.}}
\end{figure}

\begin{figure}[t]
\begin{center}
\includegraphics[width=8cm]{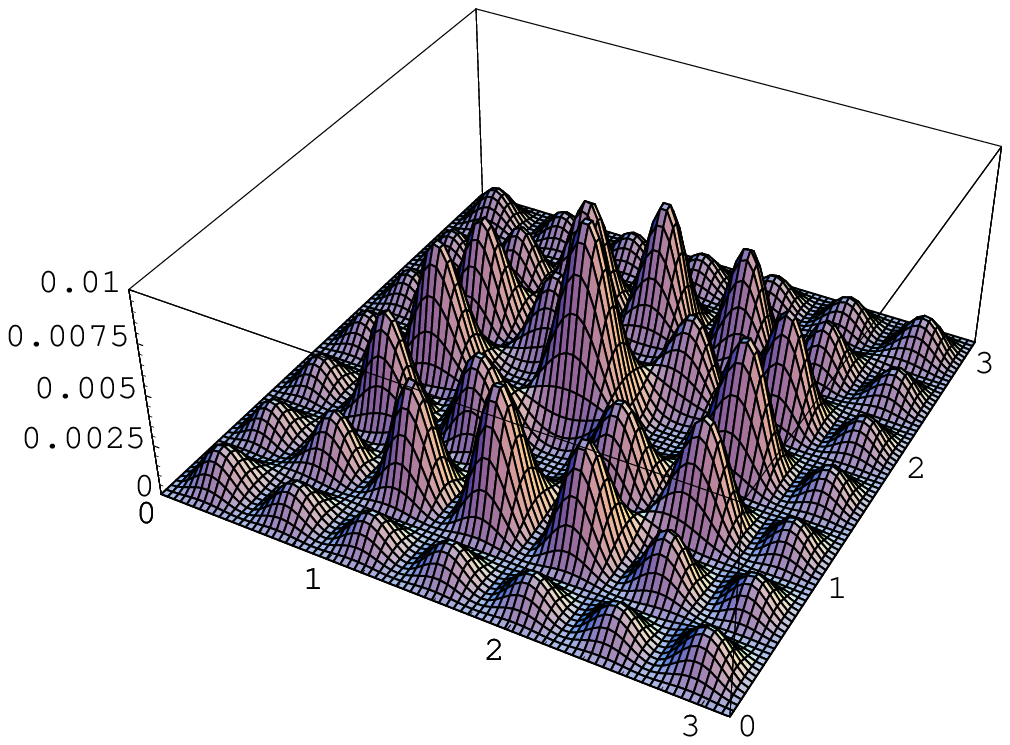}
\includegraphics[width=8cm]{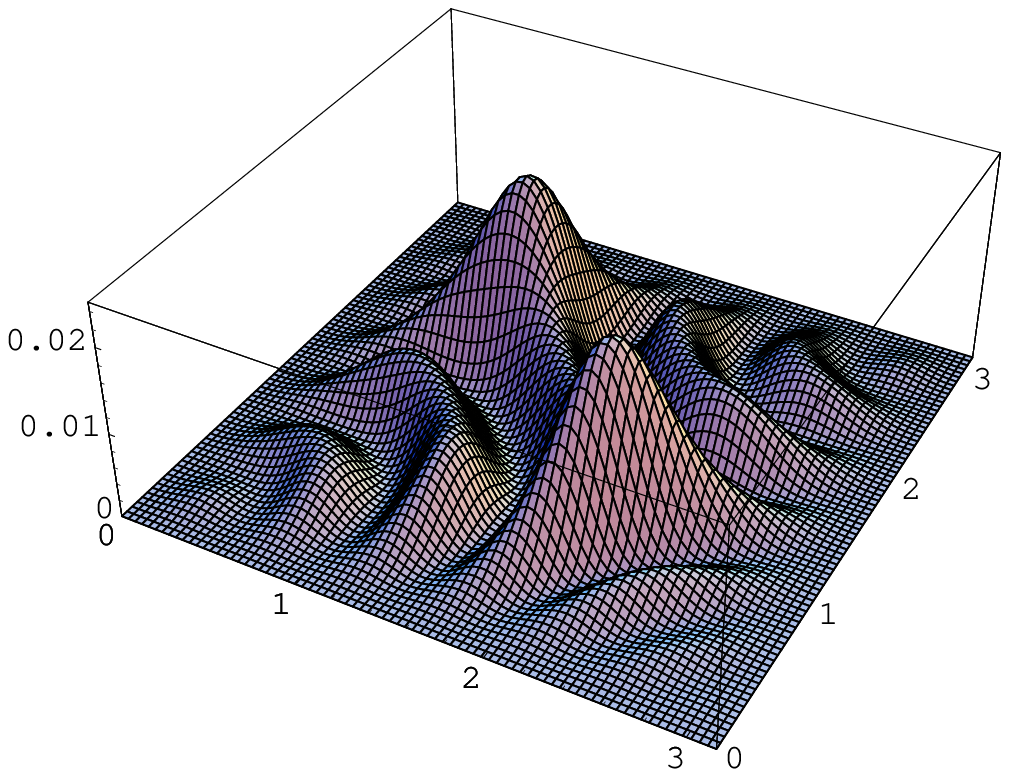}
\end{center}
\caption{\label{figtheta2}
\small{Probability distribution
$p(\theta_1,\theta_2,\varphi=\pi/2)$ and
$p(\theta_1,\theta_2,\varphi=\pi/21)$ for $j_1=j_2=j_3=3$.}}
\end{figure}

\begin{figure}[t]
\begin{center}
\includegraphics[width=8cm]{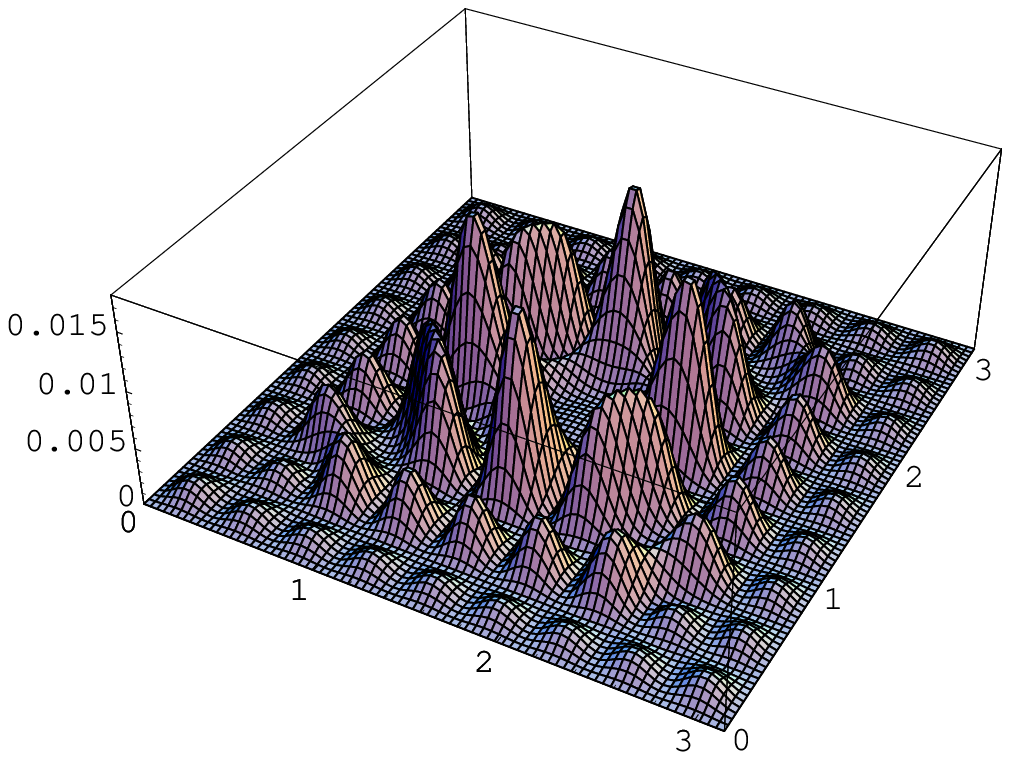}
\includegraphics[width=8cm]{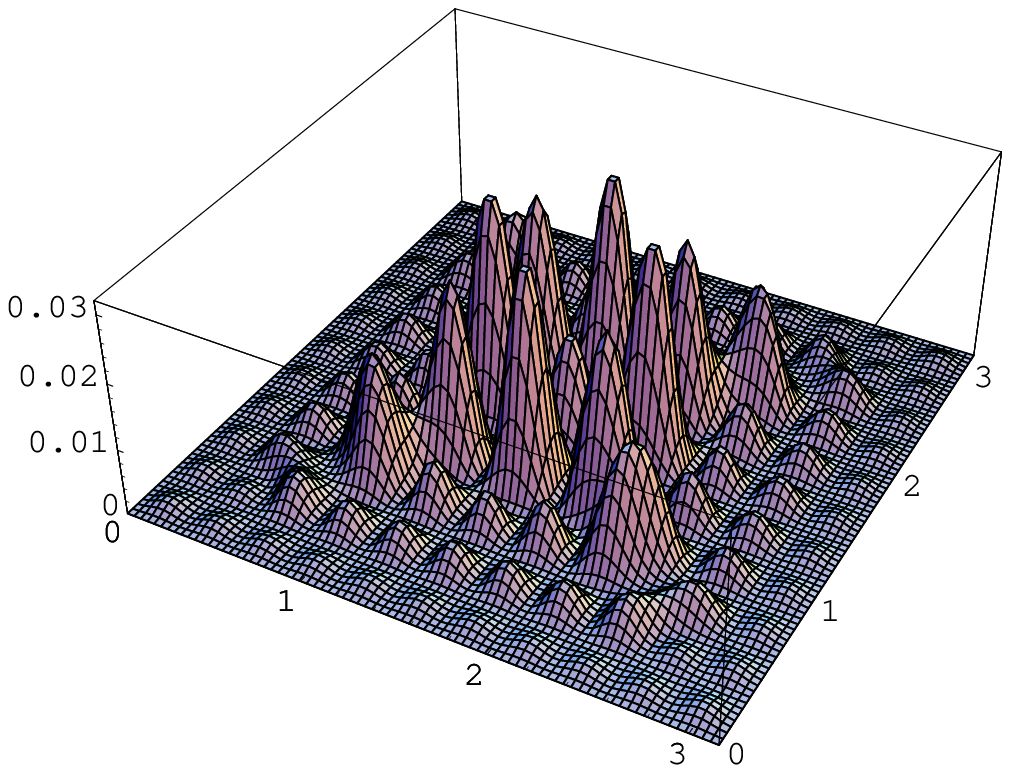}
\end{center}
\caption{\label{figtheta3}
\small{Probability distribution
$p(\theta_1,\theta_2,\varphi=\pi/2)$ for $j_1=j_2=j_3=4$
and $j_1=j_2=j_3=6$.}}
\end{figure}


\end{expl}

In general, more complicated spin networks will have more complex
probability graphs. A generic feature though is that the flat
connection $\forall e\in\Gamma,\, g_e=\Id$ will have a vanishing
probability (density). Nevertheless, when the spins $j_e$ grows to
$\infty$, it is likely that we obtain maxima arbitrarily close to
the flat configuration (like in the case of a single Wilson loop).

Let us emphasize that these are only ``kinematical" probabilities,
based on the kinematical scalar product. ``Physical" probabilities
would be based on the physical scalar product taken into account
the projection onto the kernel of the Hamiltonian constraint. The
kinematical probabilities describe the unconstrained 3d space
geometry, whereas the physical probabilities should implement the
canonical constraints imposed by general relativity on the 3d
geometry and should reflect the dynamical transition amplitudes
induced by the Hamiltonian constraint.

In the present work, we aim to understand the geometry and
structure of the quantum 3d geometry states, or spin networks, at
the kinematical level. Indeed, before implementing the Hamiltonian
constraints or the Einstein equations at the classical level, we
first have to grasp the notion of metric. Here, we would like to
understand the notion of quantum metric, before studying the
solutions to the quantum constraint which should implement quantum
gravity.

\section{Surface States and Area Renormalisation}
\label{AreaRenorm}

Let us consider a surface $\ss$ in the 3d space $\Sigma$ and a
spin network state $|\Gamma,\v{\jmath},\vio\ra$ based on the graph
$\Gamma$ considered as embedded in $\Sigma$. The surface
intersects the graph\footnote{Here we restrict ourselves to a
simpler generic situation when the surface $\ss$ does not
intersect the graph $\Gamma$ at any vertex. Considering that case
would not complicate the mathematics but only the notations.} at a
certain number of points $\ss\cap
\Gamma=\{P_1,..,P_n\}$ belonging to the graph edges $P_i\in e_i$.
We then cut the surface $\ss$ in elementary patches $\ss_i$
intersecting $\Gamma$ at the points $P_i$: $\ss=\coprod_i
\ss_i$ with $\ss_i\cap\Gamma=P_i$. The area (operator) of $\ss$ is
defined as the sum of the areas (operators) of the elementary
patches:
\be
\what{\aa}_\ss\,|\Gamma,\v{\jmath},\vio\ra
\,\equiv\,
l_P^2\sum_i S(j_{e_i})\,|\Gamma,\v{\jmath},\vio\ra.
\ee

This simple definition of the area operator of a surface $\ss$ in
LQG raises two natural issues, due to the background independence
of the theory. The first question is the definition of the surface
itself in the diffeomorphism-invariant framework of LQG. Indeed,
forgetting about the embedding of $\ss$ and $\Gamma$ in the space
manifold $\Sigma$, the surface $\ss$ is defined on the spin
network state $|\Gamma,\v{\jmath},\vio\ra$ only through its
intersections $P_i$ with the graph $\Gamma$. Starting with a
quantum geometry defined solely from the state
$|\Gamma,\v{\jmath},\vio\ra$, independently from any metric or
embedding or reference to $\Sigma$, one can define a (quantum)
surface as a set of edges $e_1,..,e_n$ of the graph $\Gamma$: the
surface is then defined\footnotemark as the union of elementary
patches $\ss_i$ transversal to each edge $e_i$. However, how can
we know that these patches are close to each over and that they do
form a smooth continuous surface? The second question is related
and concerns the semi-classical limit of such a quantum surface.
More precisely, we would like to know how the surface is folded,
i.e how the elementary patches are organized with respect to each
other. Since these elementary patches might be crumpled giving a
fractal-like structure to the surface, we  would like to be able
to coarse-grain the surface, define its macroscopic structure and
its macroscopic area.

\footnotetext{
Here, we consider the case of a fixed graph $\Gamma$. States in
LQG are actually defined as a vector in the projective limit of
the Hilbert spaces attached to all possible graphs $\Gamma$. In
some sense, a state $|\vphi\ra$ is defined as a family of
``compatible" states $\{\vphi_\Gamma\}\in\underset{\Gamma}{\times}
\hh_\Gamma$, such that $\vphi_{\Gamma_1}$ is (more or less)
identified to $\vphi_{\Gamma_2}$ through a certain natural
projection if $\Gamma_2$ is included in the graph $\Gamma_1$. A
surface (and more generally a region of space) should be defined
within this framework as ``consistent" assignments of edges on all
possible graphs $\Gamma$, in such a way that it is made compatible
with the projective structure.}

Following the preliminary ideas in \cite{ourbh}, we will propose a
framework of {\it area renormalisation} to address these issues
and we will define coarse-grain area operators to probe the
quantum structure of a surface.

\medskip

Let us fix a graph $\Gamma$ and an arbitrary quantum state defined
as the cylindrical function $\psi_\Gamma(g_e)$. Let us consider
the quantum surface defined as the set of edges $\{e_1,..,e_n\}$.
The area operator $\what{\aa}_i$ of the elementary surface $\ss_i$
transversal to the edge $e_i$ is defined as the Casimir operator
of the following $\SU(2)$ action on $\psi_\Gamma(g_e)$:
\be
(g_{e_i},g_{e\ne e_i})
\,\underset{g\in\,\SU(2)}\longrightarrow\,
(gg_{e_i},g_{e\ne e_i}).
\ee
In this simple case, it doesn't matter whether we act by left or
right multiplication. It is natural to generalize this area
operator to collections of elementary surfaces. Indeed, let us
group together the edges $e_1$ and $e_2$. We define the
coarse-grained area operator $\what{\aa}_{(12)}$ as the Casimir
operator of the diagonal $\SU(2)$ action:
\be
(g_{e_1},g_{e_2},g_{e\ne e_1,e_2})
\,\underset{g\in\,\SU(2)}\longrightarrow\,
(gg_{e_1},gg_{e_2},g_{e}).
\ee
This is easily generalized to any collection of edges. For
example, the complete coarse-grained area operator
$\what{\aa}_{1..n}$ of the surface $\ss$ is defined as the Casimir
operator of the diagonal $\SU(2)$ action:
\be
(g_{e_1},g_{e_2},..,g_{e_n},g_{e\ne e_1,..,e_n})
\,\underset{g\in\,\SU(2)}\longrightarrow\,
(gg_{e_1},gg_{e_2},..,gg_{e_n},g_{e}).
\ee
It is now necessary to distinguish the left and right
multiplication. We actually define $2^n$ different operators. Not
only we can define the two operators, $\what{\aa}_{(12)}^{(L)}$
and $\what{\aa}_{(12)}^{(R)}$, corresponding to the actions
$(g_1,g_2)\arr(gg_1,gg_2)$ and  $(g_1,g_2)\arr (g_1g,g_2g)$, but
we must also consider the two other actions $(g_1,g_2)\arr
(gg_1,g_2g^{-1})$ and $(g_1,g_2)\arr (g_1g^{-1},gg_2)$.

\begin{figure}[t]
\psfrag{j}{$j=12$}
\psfrag{j1}{$j_1=1000$}
\psfrag{j2}{$j_2=1007$}
\psfrag{S}{$\ss$}
\begin{center}
\includegraphics[width=5cm]{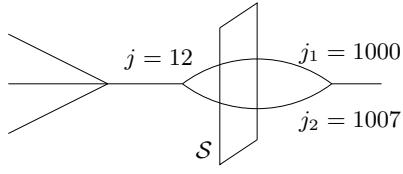}
\end{center}
\caption{\label{1000to12}
\small{
Considering a surface $\ss=\{e_1,e_2\}$ made of two spin network
edges labelled by spins $j_1=1000$ and $j_2=1007$, the microscopic
area is $S(1000)+S(1007)$ while the coarse-grained area $S(12)$ is
much smaller and takes into account the nearby intertwiner
describing the  coupling between the two representations $j_1$ and
$j_2$.}}
\end{figure}

The key point is that $\what{\aa}_{(12)}\,\psi_\Gamma\ne
\left(\what{\aa_{1}}+\what{\aa_{2}}\right)\,\psi_\Gamma$ and this
results in a non-trivial flow of the area of a surface under
coarse-graining. Let us consider the simple example of a spin
network state with the edges $e_{1,2}$ labelled with the
representations $j_1=1000$ and $j_2=1007$. The microscopic area of
the surface $\ss\equiv\{e_1,e_2\}$ is $\aa_{\rm
micro}=S(1000)+S(1007)$ in Planck unit. Now assuming that these
two edges meet at a 3-valent vertex as in fig.~\ref{1000to12} and
that the third edge is labelled with the representation $j=12$,
then the macroscopic (coarse-grained) area of the same surface
$\ss$ will be $\aa_{\rm macro}=S(12)$ which is substantially
different.

We propose to use the coarse-graining flow, or renormalisation
flow, of the surface area to probe the structure of the surface. A
possible coarse-graining of the surface $\ss=\{e_1,..,e_n\}$ is
summarized by a tree $T$ which describes how we group the edges
together: each level of the tree defines a partition of
$\{e_1,..,e_n\}$ which is coarser than the previous level, from
the initial point with individual edges to the final point when
all the edges are grouped together. The coarse-grained area
(operator) of the surface $\ss$ at each level depends on the
partition and is defined as the sum of the area operators of each
collection of edges of the given partition. Then each tree $T$
describes a possible coarse-graining of the surface $\ss$ from the
microscopic realm to the macroscopic level. Given  a state of
geometry $\psi_\Gamma$, we can analyze the renormalisation flow of
the expectation value of the area for each coarse-graining tree
$T$.

For some trees the flow may be smooth, meaning that transitions
from a level to a neighboring lead to small changes in area in
Planck units, while it might be jumpy for the other trees. We
expect that if there does not exist any tree with smooth flow,
then the set of chosen edges can not describe a true surface in
the macroscopic limit and that these edges are likely to be far
from each other in the geometry defined by the state
$\psi_\Gamma$. On the other hand, a tree with a smooth flow should
give us the right way to put the elementary surfaces together
grouping them by closest neighbors. Moreover, we expect that such
a (smooth) area flow tells us more about the properties of the
surface such as its curvature. For example, a (nearly) flat
surface should have a almost trivial constant area flow.

This proposes an explicit relationship between the geometry of a
spin network state and its renormalisation properties. This will
be investigated in more details through specific examples in
future work \cite{elisa}. A similar analysis of the volume
renormalisation flow should carry even more information about the
geometry of the quantum state $\psi_\Gamma$.

This is to be compared with theories on the lattice.
Renormalisation \`a la Wilson is defined through coarse-graining
the lattice and deriving the flow of the Hamiltonian and of the
various observables under that coarse-graining operation. Usually,
the lattice is embedded in flat space and the coarse-graining
defined through the resulting notion of nearest neighbor. In our
quantum gravity context, spin networks can be thought as a
lattice. However, there are provided without any embedding into a
space manifold. Our proposal is to consider all the possible
renormalisation flow of geometrical observables (such as the area)
and to reconstruct a ``correct" embedding by identifying the
``good" renormalisation flow(s). The whole issue is then to define
what we mean by ``good". The simplest criteria is the smoothness
of the flow. However, this should be studied in more details.

\medskip

In the case of a (pure) spin network state
$\psi_\Gamma=\,|\Gamma,\v{\jmath},\vio\ra$, we can give a simple
expression of the expectation value of a coarse-grained area
operator. Let us consider the collection of edges $\{e_1,..,e_n\}$
and the Casimir operator of the diagonal $\SU(2)$ action
$(g_{e_1},g_{e_2},..,g_{e_n})\arr (gg_{e_1},gg_{e_2},..,gg_{e_n})$
where we act at the source vertex of all edges. Considering the
action where $g$ sometimes acts at the target vertex would not
change much to the following.

\begin{figure}[t]
\psfrag{1}{$e_1$}
\psfrag{2}{$e_2$}
\psfrag{3}{$e_3$}
\psfrag{S}{$\ss$}
\psfrag{A}{$A$}
\psfrag{B}{$B$}
\begin{center}
\includegraphics[width=5cm]{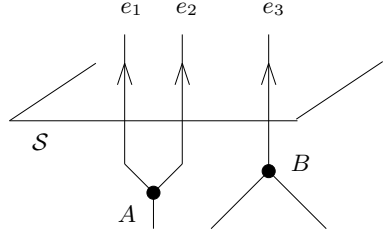}
\end{center}
\caption{\label{3edgesurf}\small{
Considering a surface $\ss=\{e_1,e_2,e_3\}$ made of three spin
network edges labelled by representations $j_{1,2,3}$, the surface
state relevant for the (coarse-grained) area expectation values is
the mixed state on the Hilbert space $V^{j_1}\otimes
V^{j_2}\otimes V^{j_3}$ defined by the density matrix
$\rho_{a_ib_i}=\overline{\ii_A}^{a_1a_2\alpha}\ii_A^{b_1b_2\alpha}
\overline{\ii_B}^{a_3\beta\gamma}\ii_B^{b_3\beta\gamma}$.
It depends on the intertwiners living at the vertices $A$ and $B$
at the source of the edges $e_{1,2,3}$.}}
\end{figure}

Using the kinematical scalar product on the spin networks defined
by the integration with the Haar measure, we show that the area
expectation value is given by:
\be
\la \what{\aa}_{(e_1,..,e_n)} \ra_\psi\,=\,
\f{\tr\,\v{J}^2\,\rho}{\tr \rho},
\ee
where $\rho$ is a density matrix describing the state of the
surface $e_1,..,e_n$ in the Hilbert space $\otimes_i V^{j_{e_i}}$.
$\rho$ is defined in terms of the intertwiners $\ii_v$ of the spin
network at the vertices where we act with $g$, i.e
$v=s(e_1),..,s(e_n)$:
\be
\rho_{a_1..a_nb_1..b_n}\,\equiv\,
\prod_i \tr\left[\overline{\ii_{v=s(e_i)}}{}^{a_i}\,\ii_{v=s(e_i)}{}^{b_i}\right],
\ee
where $a_i,b_i$ label a basis of the representation space
$V^{j_{e_i}}$. The trace is over the representation space
associated to the edges $e$ not belonging to the surface
$\{e_1,..,e_n\}$. An example is given in fig.~\ref{3edgesurf}.
$\rho$ is hermitian and generically represents a mixed state.

From the above expression for the surface state, it is clear that
the coarse-graining procedure probes the structure of the spin
network around the considered surface, for instance the
intertwiners attached to the vertex adjacent to the surface. These
intertwiners describe how the surface is folded, i.e the
"position" of the elementary patches with respect to each other.
If we were to use a more complicated scalar product between spin
network states, taking into account some dynamical effect and
propagation of the geometry degrees of freedom, it is likely that
the coarse-grained area expectation value would involve
intertwiners attached to further vertices and would depend in a
more complex way on the spin network geometry.

\medskip

We conclude this section with a simple example of area
renormalisation. We consider three edges $e_{1,2,3}$ all labelled
by the fundamental representation $j=\f12$. The representation
space $V^{1/2}$ is two-dimensional and we write the two basis
vector of maximal and minimal weights as
$|\uparrow\ra,|\downarrow\ra$. Let us consider the surface state
defined as the pure state
$\rho=|\uparrow\uparrow\downarrow\ra\la\uparrow\uparrow\downarrow|$.
Then the microscopic area in Planck unit is obviously:
$$
\la \what{\aa}_{123} \ra = 3\, S\left(\f12\right) \sim 2.598,
$$
where the numerical value is computed using the spectrum
$S(j)=\sqrt{j(j+1)}$. The coarse-grained areas, the one grouping
the edges 1,2 together and the other grouping the edges 2,3
together are easily computed\footnotemark:
$$
\la \what{\aa}_{(12)3} \ra = S(1) + S\left(\f12\right) \sim 2.280,
$$
$$
\la \what{\aa}_{1(23)} \ra = S\left(\f12\right)+\f12\left(S(1)+S\left(\f12\right)\right) \sim 2.006.
$$
\footnotetext{
We would like to decompose the space $V^{1/2}\otimes V^{1/2}$ into
irreducible representations of (the diagonal action of) $\SU(2)$.
It is well-known that $V^{1/2}\otimes V^{1/2}=V^0\oplus V^1$ with:
$$
|j=0\ra=\f1{\sqrt 2}
\left(|\uparrow\downarrow\ra-|\downarrow\uparrow\ra\right),\qquad
|j=1,m=1\ra=|\uparrow\uparrow\ra, \quad|j=1,m=0\ra=\f1{\sqrt
2}\left(|\uparrow\downarrow\ra+|\downarrow\uparrow\ra\right),\quad
|j=1,m=-1\ra=|\downarrow\downarrow\ra.
$$
}

Finally the totally coarse-grained area grouping all three edges
1,2,3 together is given by:
$$
\la \what{\aa}_{(123)} \ra = \f13S\left(\f12\right) +
\f23S\left(\f12\right) \sim 1.223.
$$
We can not really introduce a notion of smooth flow for such a
small surface with so few edges. We would need a much larger
surface, for which the state $\rho$ would inevitably be much more
complicated and analytical computations more involved
\cite{elisa}. Nevertheless, there is one state which can easily
studied. It is the pure state for $n$ edges,
$\rho=|\uparrow..\uparrow\ra\la\uparrow..\uparrow|$. Its
renormalisation flow is practically constant. If we
partition/group the set of $n$ edges into packets of size
$\alpha_i$, $\sum_i \alpha_i=n$, then the corresponding
coarse-grained area is:
$$
\la \what{\aa}_{\{\alpha_i\}} \ra=\sum_i S\left(\f{\alpha_i}
2\right).
$$
The microscopic area is $\la \what{\aa}_{\rm micro} \ra= nS(\f12)$
while the macroscopic area is $\la \what{\aa}_{\rm micro} \ra=
S(\f n2)$. If we use the simpler spectrum $\wt{S}(j)=j$, the
renormalisation is actually exactly constant.

\section{Coarse-Graining: from Bulk to Boundary}
\label{Coarse}

Considering a bounded region $A$ of a spin network state, we are
interested by the boundary state(s) on $\pp A$ induced by the
quantum state of geometry of the bulk of $A$ (or by the region
outside the region $A$). We will see the boundary state can be
considered as a coarse-graining of the geometry state of $A$ and
be constructed from partial tracing of the bulk degrees of
freedom.


\subsection{Coarse-graining: definition}
\label{coarse-a}

Let us consider quantum states based on a fixed given graph
$\Gamma$, which we assume closed and connected, and a bounded
connected region $A$ of this spin network which we define as a set
of vertices in $\Gamma$.

An edge $e$ is considered in $A$ if both its source and target
vertices are in $A$, $s(e),t(e)\in A$. We define the boundary $\pp
A$ of the region $A$ as the set of edges with one vertex in $A$
and one vertex outside $A$. For every couple of vertices $v_1,v_2$
in $A$ there exists a path of edges in $A$ linking them. Moreover,
if one cuts in two all the edges in the boundary $\pp A$, then one
cuts the graph $\Gamma$ in two connected open graphs. We will call
$\Gamma_A$ the connected open graph with all the vertices in $A$
and ending at the boundary $\pp A$. Finally, we call $\breve{A}$
the exterior of $A$ i.e the set of vertices and edges who do not
belong to $A$ and its boundary $\pp A$.

For the sake of the simplicity of the notations, we assume in the
following that all the edges $e\in \pp A$ on the boundary are
identically oriented outward, with $s(e)\in A$ and $t(e)\notin A$.

\begin{figure}[t]
\begin{center}
\includegraphics[width=8cm]{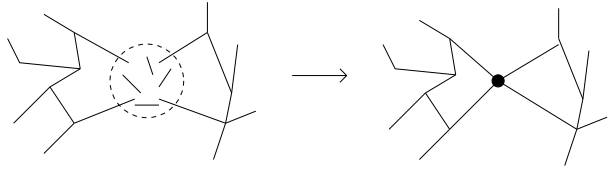}
\end{center}
\caption{\label{GammaA}
\small{
Coarse-graining the region $A$ to a single vertex: from the graph
$\Gamma$ to the reduced graph $\Gamma[A]$. }}
\end{figure}

The boundary state of $A$  is the state of the edges $e\in \pp A$
irrespectively of the details of the structure of the graph
$\Gamma_A$ inside the region $A$. We can therefore think about it
as the coarse-grained state of $A$ once we coarse-grain the whole
interior graph of $A$ to a single vertex. Let us thus define the
graph $\Gamma[A]$ where we have removes the whole graph $\Gamma_A$
from $\Gamma$ and replaced it by a single vertex, as in
Fig.\ref{GammaA}.
We then define the following coarse-graining map.
\begin{defi}
\label{def1}
We define a coarse-graining map $\Theta:H_\Gamma \arr
H_{\Gamma[A]}$ taking cylindrical functions defined on the full
graph $\Gamma$ to cylindrical functions defined on the reduced
graph $\Gamma[A]$:
\be
(\Theta\,\psi)(\{g_e, e\in\Gamma[A]\})
\,\equiv\,
\int_{\SU(2)}dg\,
\psi(\{g_{e\in \breve{A}},gg_{e\in\pp A},g^{(0)}_{e\in A}\}).
\ee
The integration over $g$ is to ensure the proper gauge invariance
of the resulting functional. The map $\Theta$ actually depends on
the group elements $g^{(0)}_{e\in A}\in\SU(2)$, which need to be
fixed. We call them the coarse-graining data. We call the trivial
or canonical coarse-graining map the one defined by $g^{(0)}_{e\in
A}=\Id$. For this canonical  choice, the integration over $\SU(2)$
is irrelevant and we can drop it.
\end{defi}
Due to the gauge invariance of the initial functional $\psi$, the
coarse-graining data $g^{(0)}_{e\in A}$ is actually defined up to
$\SU(2)$ actions at each vertex of $A$.
As a result,  the coarse-graining data lives in the quotient space
$X_A\,\equiv\,\SU(2)^{E_A}/\SU(2)^{W_A+1}$, where $W_A$ is the
number of vertices in $A$ who do not have any edge on the
boundary.

\medskip

It is useful to look at this coarse-graining procedure in the case
of a (pure) spin network state. The spin network state labels each
edge $e$ of the graph with a (irreducible) representation $j_e$.
The state of the bulk of the region $A$ is given by the
intertwiners attached to the vertices of $A$, and lives in the
Hilbert space $\hh_A\,\equiv\,\otimes_{v\in A}\Int_v$. The state
of the boundary $\pp A$ lives in the tensor product $\otimes_{e\in
\pp A} V^{j_e}$. We require it to be $\SU(2)$ invariant so that
the boundary Hilbert space is $\hh_{\pp A} \, \equiv \,
\Int(\otimes_{e\in \pp A} j_e)$. The coarse-graining is a map from
$\hh_A$ to $\hh_{\pp A}$: we glue the intertwiners inside $A$
together in order to obtain an intertwiner solely between the
edges on the boundary of $A$. This procedure is equivalent to the
construction described above and effectively reduces the whole
interior graph of $A$ to a single vertex.

More precisely, given some fixed holonomies along the edges of
$A$, we can define the coarse-grained intertwiner as follow.
\begin{defi}\label{def2}
Given the coarse-graining data $\{g^{(0)}_{e\in A}\}$, and
intertwiners $\ii_{v\in A}\in \hh_A$, we define the following
coarse-grained intertwiner in $\hh_{\pp A}$:
\be
\ii_A[g^{(0)}_{e\in A}]\,\equiv\,
\int_{\SU(2)}dg\, \tr_{e\in A}\,\left[
\bigotimes_{e\in \pp A} D^{j_e}(g)
\bigotimes_{e\in A} D^{j_e}(g^{(0)}_{e})
\bigotimes_{v\in A}\ii_v.
\right]
\ee
The integration over $\SU(2)$ is to impose $\SU(2)$ invariance and
obtain an intertwiner at the end of the day. The canonical choice
$\{g^{(0)}_{e\in A}=\Id\}$ amounts to straightforwardly gluing the
intertwiners. In that case, we automatically respect gauge
invariance and we obtain an intertwiner without having to
integrate over $\SU(2)$.
\end{defi}
More details on the motivation for this definition of
coarse-graining from the point of view of partial tracing and the
uniqueness of the procedure are given in appendix.

Due to the gauge invariance, the coarse-grained intertwiner
actually depends only on the orbit of $\{g^{(0)}_{e\in A}\}$ under
$\SU(2)$ transformations: it is defined in terms of $x\in
X_A=\SU(2)^{E_A}/\SU(2)^{W_A+1}$.

\begin{expl}
Let us explain the simple example of a 4-punctured boundary
$V_{j_1}\otimes V_{j_2}\arr V_{j_3}\otimes V_{j_4}$  with the
internal graph being made of two 3-valent vertices and a single
internal edge labelled with a fixed spin $j$. The two intertwiners
attached to the two 3-valent vertices are unique and given by the
usual Clebsh-Gordan coefficients. Coarse-graining means assigning
a group element to the internal edge and integrating over an orbit
under the adjoint action $Ad(\SU(2))$. An orbit in
$\SU(2)/Ad(\SU(2))$ is defined by an angle $\theta\in ]-\pi,\pi]$
and is the equivalence class of the group element $h_\theta\equiv
\exp(i\theta J_z)$. The corresponding coarse-grained intertwiner
$\ii^\theta:V_{j_1}\otimes V_{j_2}\arr V_{j_3}\otimes V_{j_4}$
becomes
\be
\ii^\theta_{m_1m_2m_3m_4}=\int_{\SU(2)}dg\,
C^{j_3j_4j}_{m_3m_4n}D^j_{mn}(gh_\theta
g^{-1})C^{j_1j_2j}_{m_1m_2m}.
\ee
We can evaluate exactly the integral over $g$:
\be
\int_{\SU(2)}dg\,D^j_{mn}(gh_\theta g^{-1})=
\delta_{mn}\f{\chi^j(\theta)}{2j+1}.
\ee
We obtain that $\ii^\theta=\ii^0\,\chi^j(\theta)/(2j+1)$ always
leads to the same intertwiner defined at $\theta=0$ up to a
normalization factor: the coarse-graining procedure is trivial in
this simple setting. If the Clebsh-Gordan coefficients are
normalized, then the norm of $\ii^{\theta=0}$ is 1. As we will see
later, the reason why the coarse-graining procedure is trivial is
that the topology of the interior graph (here a single edge) is
trivial. Coarse-graining will give a non-trivial result only when
the interior graph has a non-trivial topology and contains at
least one loop.
\end{expl}

\medskip

We now would like to extend these intertwiner definitions from a
(pure) spin network state to an arbitrary cylindrical function. We
need to decompose the cylindrical function onto the spin network
basis, coarse-grain each spin network component separately, and
then sum the coarse-grained components back to obtain the
coarse-grained cylindrical function. This is equivalent to the
definition def.\ref{def1} of coarse-graining, since the procedure
is a linear functional only involving integrating over part of the
degrees of freedom of the wave-function (the holonomies on the
edges of $A$). Nevertheless, we ought to consider the same
coarse-graining data for all the spin network components of the
initial functional. However, it might happen that it seems more
suited to consider different data for each component, depending on
the structure of each spin network state.


\subsection{The most probable coarse-graining?} \label{probcoarse}

The natural issue  following the previous definition is how to
choose the coarse-graining data. Indeed a priori, the
coarse-grained state is a mixed state reflecting all the possible
coarse-grainings given by the density matrix:
\be
\rho_A\equiv \int_{X_A} d\mu(x)\,
|\ii_A[x]\ra\la \ii_A[x]|.
\ee
A priori the measure $\mu(x)$ is  simply the Haar measure on
$X_A=\SU(2)^{E_A}/\SU(2)^{W_A+1}$. This density matrix comes
explicitly from tracing out the holonomies along the edges within
the region $A$.

A first remark is that this density matrix is in general not the
totally mixed state $\Id_{\pp A}/{\rm dim}\hh_{\pp_A}$. Therefore,
our coarse-graining procedure does not erase all the information
about the internal state of the region $A$. Obtaining the totally
mixed state by averaging over all possible ways of coarse-graining
would thus require a special quantum state. We expect this to be
related to quantum black holes \cite{ourbh}.

The second remark is that the coarse-grained procedure does not
provide us with normalized intertwiners $ ||\ii_A[x]||\ne1$.
Following the interpretation of the spin network functional as a
wave function defining a probability amplitude on the holonomies,
this norm $||\ii_A[x]||^2$ actually provides us with a measure of
the probability on the space of possible coarse-grainings $X_A$.
Indeed, the coarse grained intertwiner $\ii_A[x]$, or equivalently
the coarse grained state defined in def.\ref{def1}, are defined
simply by evaluating the initial spin network on the holonomies
$x=\{g^{(0)}_{e\in A}\}$. This means that, although it seems more
natural to coarse-grain using the canonical choice $x=\Id$, that
might not be the most probable holonomies inside the region $A$.

To summarize the situation, let us assume that we work with
normalized spin network functionals i.e that all the intertwiners
$\ii_v$ are normalized, $||\ii_v||=1$. Now the density matrix
describing the coarse-grained geometry of the region $A$ is:
$$
\rho_A
=\int_{X_A} d\mu(x)\, |\ii_A[x]\ra\la \ii_A[x]|
=\int_{X_A} ||\ii_A[x]||^2\,d\mu(x)\, |\wt{\ii}_A[x]\ra\la
\wt{\ii}_A[x]|,
$$
where $\wt{\ii}_A[x]$ is now the {\it normalized} intertwiner
describing the geometry of the region $A$ reduced to a single
vertex. Clearly the probability distribution on this intertwiner
is given by the measure $||\ii_A[x]||^2\,d\mu(x)$.

We can finally generalize these considerations to an arbitrary
cylindrical function. The coarse-grained state will similarly be
described by the density matrix,
$$
\rho_A(\psi)\,\equiv\,
\int dx |\Theta_x\psi\ra\la \Theta_x\psi|,
$$
and the most probable coarse-grainings will be defined as the
maxima of the norm of the coarse-grained functional
$||\Theta_x\psi||^2$.

\subs{Coarse-graining data, gauge fixing and non-trivial topology}

To understand the structure and relevance (or irrelevance) of the
coarse-graining data, it is useful to explicitly gauge-fix the
initial functional $\vphi$. This is he simplest way to see how the
whole graph inside $A$ gets reduced to a single vertex.

Indeed, following \cite{noncompact}, we choose a (reference)
vertex $v_0$ in $A$ and a maximal tree $T$. $T$ is a set of edges
in $A$ going through all vertices of $A$ without ever making any
loop. It has $|T|=V_A-1$ elements. If $|T|=E_A$, this means that
the interior graph $\Gamma_A$ has no loop and has a trivial
topology. $E_A-|T|$ counts the number of non-trivial loops of
$\Gamma_A$. We know that we can use the gauge invariance to fix to
the identity the group elements on the edges belonging to the
tree, $g_{e\in T}=\Id$,
$$
\psi(\{g_{e\in \breve{A}},g_{e\in\pp A},g_{e\in A}\})
\,=\,
\psi(\{g_{e\in \breve{A}},H_eg_{e\in\pp A},\Id_{e\in T},G_{e\in A\setminus
T}\}).
$$
To define $G$ and $H$, we define for every vertex $v\in A$ the
unique path $\Ll_v$ along $T$ from $v_0$ to $v$. Then we define
the group element:
$$
H_v\,\equiv\,\overrightarrow{\prod_{e\in
\Ll_{v}}} g_{e}.
$$
Finally, $H_{e\in \pp A}\,\equiv\, H_{s(e)}$ and $G_{e\in
A\setminus T}\,\equiv\, H_{s(e)}g_eH_{t(e)}^{-1}$.

The $H_{e\in \pp A}$ can be considered as irrelevant for the
coarse-graining since that they can be re-absorbed in the
redefinition of the holonomies $g_{e\in\pp A}$ on the boundary of
$A$. On the other hand, the group elements $G_{e\in A\setminus T}$
can not be forgotten and actually represent the holonomies around
each (non-contractible) loop of the interior graph $\Gamma_A$.
They therefore reflect the non-trivial topology\footnotemark of
the quantum state of geometry of $A$. \footnotetext{Let us point
out that the link between the topology of the graph underlying the
spin network state with the topology of the actual manifold in a
semi-classical limit is not clear. See \cite{flo} for example.}

\begin{figure}[t]
\psfrag{j1}{$j_1$}
\psfrag{j2}{$j_2$}
\psfrag{ja}{$j_a$}
\psfrag{jb}{$j_b$}
\begin{center}
\includegraphics[width=6cm]{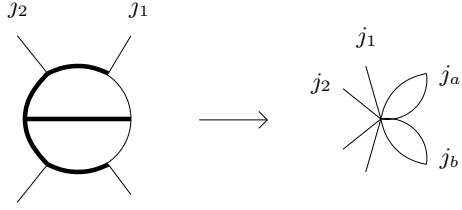}
\end{center}
\caption{\label{LoopToPetal}
\small{
Gauge-Fixing of a region with two loops: the graph has $|\pp A|=4$
external edges, $E_A=7$ internal edges, $V_A=6$ vertices,
$L_A=E_A-V_A+1=2$ loops. We gauge fix along the tree $T$ in bold
and contract the whole internal graph to a single vertex. That
vertex defines an intertwiner $V^{j_1}\otimes\cdots\otimes
V^{j_4}\otimes (V^{j_a})^{\otimes 2}\otimes (V^{j_b})^{\otimes
2}\arr\C$. The coarse-graining data is then the holonomies around
the two loops $\Ll_a,\Ll_b$. }}
\end{figure}

Having set the maximal number of holonomies to the identity inside
$A$, we have effectively reduced the whole graph $\Gamma_A$ to a
single vertex (the vertex $v_0$) with $n$ open edges (with $n=|\pp
A|$ is the number of edges in the boundary) and $L_A$ loops. The
number of loops is simply $L_A=E_A-V_A+1$. See
fig.\ref{LoopToPetal} for an example with two loops.
Contracting all the vertices of $A$ to a single one, and gluing
all the corresponding intertwiners together putting the identity
on the edges of the tree $T$, lead a single intertwiner
$\ii:V^{\otimes n}\otimes V^{\otimes 2L_A}\arr \C$. Here we write
$V$ to refer to which ever representation $V^{j}$ lives on the
considered edge (or loop). The region $A$ is then described by the
contraction of this single intertwiner with the holonomies
$G_{e\in A\setminus T}$ along the loops. Labelling the loops by
$i=1..L_A$ and noting the corresponding holonomies $G_i$, the
functional reads:
\be
\ii_{\alpha_1..\alpha_{n}\,a_1b_1..a_{L_A}b_{L_A}}\,\prod_i^{L_A}D_{a_ib_i}(G_i).
\ee
Coarse-graining this structure amounts to reduce the whole graph
to a single vertex without any loop.  From this point of view, it
is clear that a non-trivial coarse-graining of the region $A$
corresponds to a non-trivial topology of the graph $\Gamma_A$
inside the region $A$. The relevant coarse-graining data is an
element of the group $\SU(2)^{L_A}=\SU(2)^{E_A-V_A+1}$ i.e one
holonomy for each loop\footnotemark in $A$. When $E_A=V_A-1$,
there is no loop in $A$ and we are left with the trivial
coarse-graining setting all holonomies on the edges in $A$ to the
identity. Coarse-graining thus amounts to erase any non-trivial
topology of the graph of the geometry state of $A$.

\footnotetext{
We can remark that if the functional initial $\vphi$ does not
actually depend on a group element $g_{e\in A\setminus T}$, it
corresponds to having a spin network state with that edge labelled
with the trivial representation $j=0$. Such an edge labelled with
$j=0$ is equivalent to removing that edge in LQG. Removing that
edge would actually destroy the corresponding loop.}

We coarse-grain by averaging over the loop holonomies with an
arbitrary function $\Phi(G_1,..,G_{L_A})\in L_2(\SU(2)^{L_A})$:
\be
\int \prod_i dG_i\, \Phi(G_1,..,G_{L_A})
\ii_{\alpha_1..\alpha_{n}\,a_1b_1..a_{L_A}b_{L_A}}\,\prod_i^{L_A}D_{a_ib_i}(G_i).
\ee
We require that the result after integration is still an
intertwiner $V^{\otimes n}\arr\C$, i.e invariant under (the
diagonal action of) $\SU(2)$ (on $\alpha_1..\alpha_{n}$). This
imposes that $\Phi(G_1,..,G_{L_A})$ be invariant under the adjoint
action of $\SU(2)$,
$$
\Phi(G_1,..,G_{L_A})=\Phi(gG_1g^{-1},..,gG_{L_A}g^{-1}).
$$
This means that $\Phi$ is a spin network functional on the flower
graph with $L_A$ petals -see fig.\ref{LoopToPetal} for an example
with a two-petal flower.

Now, one could look for coarse-graining data which maximizes the
(norm of the) coarse-grained intertwiner, either directly as (an
orbit) $Ad(\SU(2)).(G_1^{(0)},..,G_{L_A}^{(0)})$, or equivalently
as a $Ad(\SU(2))$-invariant functional $\Phi$.  If we do not
consider $Ad(\SU(2))$-invariant coarse-graining data, we will not
obtain an intertwiner as result of the coarse-graining, but more
generally a vector in the tensor product $V^{\otimes n}$.

An alternative to imposing by hand to obtain an intertwiner after
coarse-graining would to look for maxima among all coarse-graining
data $(G_1^{(0)},..,G_{L_A}^{(0)})$, possibly non-invariant. If
the most probable holonomies along the loops are actually the
identity (or at least very close to it), then we get simply set
them to $\Id$ in the coarse-graining procedure and erase the
non-trivial topology of the graph. If the probability is actually
peaked around non-trivial values of the holonomies, this could
mean that the non-trivial topology of the graph truly reflect a
non-trivial topology of the region $A$ and that we should not
neglect the contribution/couplings of the loops to the
intertwiner. In that case, our coarse-graining procedure, which
erases the non-trivial topology, might not reflect the true
geometry of the region $A$.

\medskip

At the end of the day, we have seen how the coarse-graining
procedure can be understood as ``erasing" the non-trivial topology
of the region $A$ (more precisely, the topology of the graph
underlying the quantum state of geometry of the region $A$)
through integrating over the degrees of freedom attached to the
internal edges of the region $A$.

\subsection{Inside state vs. outside state}

Considering a pure spin network state $|\Gamma,\v{\jmath},\vio\ra$
and given a region $A$ of that spin network and an assignment
$x\in X_A$ of groups elements to each internal edge, we have seen
to define a boundary state $\ii_A(x)\in\hh_{\pp A}$. Since we are
studying a spin network state based on a closed graph, the
exterior of $A$ defines a (complementary) region of the spin
network with the same boundary. Defining the quotient space
$Y_A\equiv X_{\breve{A}}$ of all possible coarse-grainings of the
exterior of $A$, each element $y\in Y_A$ induces a boundary state
$\ii_A^{\ext}(y)\in\hh_{\pp A}$, which is a priori arbitrarily
different from the states $\ii_A^{\inte}(x)$ obtained by
coarse-graining the interior of $A$.
Let us point out that $\ii_A^{\ext}(y)$ is actually
in $\overline{\hh_{\pp A}}$ (because of the reverse orientation of
the edges), which is nevertheless isomorphic to $\hh_{\pp A}$
since we work with $\SU(2)$.

At the intuitive level, one can see the boundary state
$\ii_A^{\inte}(x)$ as the state of $\pp A$ induced by the
(quantum) metric inside $A$ while $\ii_A^{\ext}(y)$ is the
boundary state induced by the outside (quantum) metric. Then we
can look for transition functions (or transition amplitudes)
allowing to go from $\ii_A^{\inte}(x)$  to $\ii_A^{\ext}(y)$. More
precisely, we would like to identify a function
$\vphi_{x,y}(\{g_e,e\in \pp A\})$ such that:
\be
\ii_A^{\ext}(y)=\int_{\SU(2)^{|\pp A|}} \prod_{e\in\pp A}dg_e\,
\vphi_{x,y}(\{g_e,e\in \pp A\}) \bigotimes_{e\in\pp
A}D^{j_e}(g_e){\mathbf{.}}\,\ii_A^{\inte}(x).
\ee
This makes sense since $\ii_A^{\inte}(x)\in \hh_{\pp
A}=\hh^0(\bigotimes_{e\in\pp A} V_{j_e})$ lives in the tensor
product $\bigotimes_{e\in\pp A} V_{j_e}$. Actually to be rigorous,
we should write $\overline{\ii_A^{\ext}}(y)$.

Such a transition amplitude  $\vphi_{x,y}(\{g_e,e\in
\pp_A\})$ would reflect the curvature at the level of the surface
$\pp A$, between the inside and the outside of $A$. Indeed it
describes the parallel transport when crossing the surface $\pp
A$.

A solution is actually given by the spin network functional itself
$\phi_{(\Gamma,j_e,\ii_v)}[g]$. Indeed a straightforward
calculation gives that a suitable function for $x=\{G_e, e\in A\}$
and $y=\{G_e, e\in \breve{A}\}$ is defined as:
\be
\vphi_{x,y}(\{g_e,e\in \pp A\})\equiv \f{1}{\|\ii_A^{\inte}(x)\|^2}\,
\overline{\phi}\,(g_{e\in\pp A}, G_{e\in A}, G_{e\in \breve{A}}).
\ee
Since the spin network functional is actually the wave-function
defining the probability amplitude for the connection and
holonomies, this result fits with the intuition that the
``transition function" $\vphi_{x,y}(\{g_e,e\in \pp_A\})$ defines
the probability amplitude for the curvature at the level of the
surface $\pp A$.





\subs{Coarse-graining of the Hilbert space: counting the degrees of freedom}

We have seen previously that the coarse-graining of a region of a
spin network is non-trivial only when the topology of the graph is
non-trivial. We propose to look at this from the perspective of
the dimension of the Hilbert space and see how the number of
degrees of freedom get coarse-grained. More precisely, we
investigate the dimensionality of the Hilbert space of bulk states
once the structure of the boundary is fixed.

Considering the bounded region $A$ of the spin network, we look at
the topology of the open graph $\Gamma_A$ inside $A$. If the graph
$\Gamma_A$ is a single vertex, then the Hilbert space is simply
the space of intertwiners between the representations $V^{j_e}$
attached to the edges on the boundary $e\in \pp A$:
$$
\hh^A_{{\rm vertex}}\,=\,\Int(\bigotimes_{e\in\pp A}V^{j_e}).
$$
If $\Gamma_A$ is more complicated and consists of many vertices
and internal edges, a basis of the Hilbert space is given by a
choice of representation labels for the internal edges $e\in A$
and of (orthogonal) intertwiner states for the internal vertices.
This existence of intertwiners at the vertices imposes some
restrictions on the possible representation labels (such as the
triangular inequalities for trivalent vertices).

When $\Gamma_A$ has a trivial topology, i.e has no loops, then the
Hilbert space is actually still isomorphic to the same intertwiner
space:
$$
\hh^A_{{\rm trivial}}\,=\,\Int(\bigotimes_{e\in\pp A}V^{j_e}).
$$
Using a tree decomposition of a vertex is actually a standard way
of constructing a basis of the intertwiner space.

As soon as $\Gamma_A$ has a non-trivial topology, i.e contains
loops, the Hilbert space is a priori infinite-dimensional. Indeed,
one can attach to each loop $\Ll$ with an un-constrained
representation label $j_\Ll$. This is similar to the freedom of
associating an arbitrary holonomy along each loop in the previous
coarse-graining procedure. Since $j_\Ll$ is an arbitrary integer,
the resulting Hilbert space has an infinite dimension. Thus one
needs to gauge fix this freedom and fix the representation $j_\Ll$
-the flux around the loop- for each loop. Then the Hilbert space
becomes finite-dimensional but still has a larger dimension than
the intertwiner space.

\medskip

Let us look at this in details in a simple example. We consider a
region $A$ whose boundary comprises of $2n$ edges labelled with
the fundamental representation $V^{1/2}$. It is rather
straightforward to show that (see \cite{ourbh} for example):
\be
(V^{1/2})^{\otimes 2n}\,=\,
\bigoplus_{k=0}^n\, (C^{n+k}_{2n}-C^{n+k+1}_{2n})\, V^k,
\ee
where the $C$'s are the binomial coefficients. The dimension of
the intertwiner space is the degeneracy of the spin-0 space:
\be
\dim \hh^{(2n)}_{{\rm trivial}}\,=\,
C^{n}_{2n}-C^{n+1}_{2n}=\f{1}{n+1}C^{n}_{2n}.
\ee

\begin{figure}[t]
\psfrag{h}{$\f12$}
\psfrag{j}{$j$}
\begin{center}
\includegraphics[width=5.5cm]{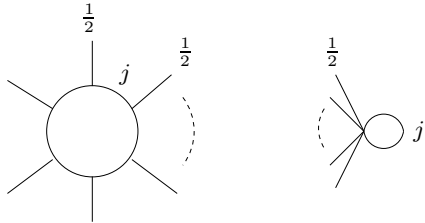}
\end{center}
\caption{\label{oneloop}
\small{
The two examples of 1-loop spin networks, with $2n$ external edges
and the loop labelled by the representation $j$. }}
\end{figure}

Let us now look at the one-loop case. We consider a loop to which
the $2n$ edges are attached, as shown in Fig.\ref{oneloop}, so
that all vertices are 3-valent. Since all vertices are 3-valent,
there is no freedom in the choice of intertwiners. All the freedom
resides in the choice of representation label for the (internal)
edges on the loop. Let us consider one of these edge. We are free
to label it with an arbitrary representation $j\in\N$. Indeed, one
can then find consistent labels on the remaining internal edges,
for example alternating $j,j+\f12,j,j+\f12,j,\dots$ on the $2n$
edges on the loop. Since $j$ has no constraint, we conclude that
the one-loop Hilbert space has a priori an infinite dimension.
However, we can gauge fix this infinity.

Indeed, starting with a given consistent labelling $\{j_e,
e\in\Ll\}$ of the edges of the loop, the freedom is actually to
add a spin $k$to all the labels. Then the labelling $\{(j_e+k),
e\in\Ll\}$ provides a new consistent labelling and thus a new
orthogonal spin network. To gauge fix this freedom, we need to fix
the spin $j_e$ on a {\it single} edge on the loop (we do {\it not}
fix the labels on all the edges on the loop, that would be too
much).

This action by arbitrary shift of the spins $j_e\arr j_e+k$ along
the loop is actually equivalent to the action on the spin network
state of the holonomy operator along that loop. Indeed acting with
the operator multiplication by the character $\chi^k(g)$ of the
$k$-representation amounts to tensor all the representations
$V^{j_e}$ along the loop by the representation $V^k$. Therefore,
we are gauge fixing the action of the (gauge invariant) holonomy
operators on loops inside the region $A$. Since the action of the
Hamiltonian constraints is intimately intertwined to the holonomy
operators (see for example \cite{lqg}), at the end of the day, our
gauge fixing could be related to gauge fixing the Hamiltonian and
therefore counting the number of physical degrees of freedom.
Actually, we already know that the gauge fixings of the holonomy
operators and of the Hamiltonian constraints are closely related
for a topological BF-type theory. However, we expect the situation
to be more subtle in 3+1 gravity.

Back to our dimension counting, we fix the spin to $j$ on a given
edge on the loop and we know count the exact number of consistent
labelling of the other internal edges. We notice that from one
edge to the next, we can only move from a representation $k$ to
$k\pm \f12$. Then along the loop, we start at $j$ and we move by
$\pm\f12$ steps to finally come back to $j$. Thus the dimension of
the one-loop Hilbert space is given by the number of returns (to
0) of a random walk after $2n$ steps:
\be
\dim \hh^{(2n)}_{1-{\rm loop}}\,=\,
C^{n}_{2n}.
\ee
This actually works only if $j\ge n/2$, else the random walk could
possibly reach the trivial representation $k=0$. Since the
representation labels are positive, the dimension of the Hilbert
space would be smaller.

Another possible one-loop configuration is a single vertex with
the $2n$ edges sticking out and one loop (or petal) going from the
vertex to itself, as shown on Fig.\ref{oneloop}. It is clear that
the loop can carry whatever representation $j\in\N$  we want. Once
again, to avoid an infinite number of degrees of freedom, we fix
the spin $j$ on the loop. We then need to count the number of
intertwiners $Int(V^j\otimes V^j\otimes (V^{1/2})^{\otimes 2n})$.
Taking into account that $V^j\otimes V^j=\oplus_{k=0}^{2j} V^k$,
the dimension of the (extended) intertwiner space is
straightforward to compute:
\be
\dim \hh^{(2n)}_{1-{\rm loop}}\,=\,
\sum_{k=0}^{\min (2j,n)} \left(C^{n+k}_{2n}-C^{n+k+1}_{2n}\right)
\,=\,C^n_{2n}-C^{n+\min (2j,n)+1}_{2n},
\ee
which is always larger than $C^n_{2n}-C^{n+1}_{2n}$. As soon as
$2j\ge n$, we recover the previous result, $\dim \hh^{(2n)}_{{\rm
1-loop}}=C^n_{2n}$.

We can then compare the number of degrees of freedom for a trivial
graph topology to the one-loop case:
$$
\log\dim \hh^{(2n)}_{{\rm trivial}} \sim 2n\log 2 -\f32\log n
\quad < \quad
\log\dim \hh^{(2n)}_{1-{\rm loop}} \sim 2n\log 2 -\f12\log n.
$$
We further expect the number of degrees of freedom to increase
with the number of loops on the graph inside the considered region
$A$. These extra degrees of freedom are erased as we coarse-grain
the region $A$ and fix the holonomies around the non-trivial loops
of the graph to finally reduce the whole region $A$ to a single
vertex.

\section{Entanglement within the Spin Network}
\label{Entanglement}

\subs{Correlations and entanglement: definitions }

Let us start by recalling definition of correlations and
entanglement between two quantum systems \cite{info}.

\begin{defi}
\label{DefEntan}
Let us consider a possibly mixed state on $\hh=\hh_A\otimes\hh_B$
defined by the density matrix $\rho$, $\rho\dag=\rho$, $\rho\ge
0$, $\tr\rho=1$. We define the reduced density matrices obtained
by partial tracing:
$$
\rho_A=\tr_B\rho,\quad \rho_B=\tr_A\rho.
$$
The entropy of the state $\rho$ is $S[\rho]=-\tr\rho\log\rho$.
Then we define the {\it quantum mutual information} as:
$$
I(A|B,\rho)\,\equiv\, S[\rho_A]+S[\rho_B]-S[\rho].
$$
Note that when we have a pure state $\rho=|\psi\ra\la\psi|$, then
the entropies of reduced density matrices are equal,
$S[\rho_A]=S[\rho_B]$.
\end{defi}

\begin{defi}
For a pure state $\rho=|\psi\ra\la\psi|$, we define the {\it
entanglement} between $A$ and $B$ as the entropy of the reduced
density matrices:
\be
\Ee(A|B,|\psi\ra\la\psi|)
\,\equiv\,
S[\rho_A]=S[\rho_B].
\ee
Note that since $S[\rho]=0$, we have $I(A|B)=2\Ee(A|B)$.

For a generic mixed state $\rho$, we define the entanglement (of
formation) as the minimal entanglement over all possible pure
state decompositions of $\rho$. More precisely, we diagonalize
$\rho=\sum_i \om_i |\psi_i\ra\la\psi_i|$. This decomposition is
not necessarily unique and we define:
\be
\Ee(A|B,\rho)\,\equiv\,
\min_{\{|\psi_i\ra\}}\,
\sum_i \om_i \,\Ee(A|B,|\psi_i\ra\la\psi_i|).
\ee
The minimization is necessary in order to obtain a meaningful
result.
\end{defi}

Other measures of entanglement are used in the literature and
reflect different aspects of entanglement manipulations. The
quantum mutual information $I(A|B)$ quantifies the total amount of
(classical and quantum) correlations between $A$ and $B$. The
entanglement $\Ee(A|B)$ defines a measure of purely quantum
correlations between the two systems $A$ and $B$. One can quantify
the classical amount of correlations as the difference
$I(A|B)-\Ee(A|B)$.

\subsection{Correlations between two parts of the spin network}

Let us consider a quantum state $\psi_\Gamma$ defined on the
(closed and connected) graph $\Gamma$, and two small\footnote{We
mean that the number of vertices in $A$ and $B$ is small compared
to the total number of vertices of the graph.} regions $A$ and $B$
of that graph. Let us first assume that $A$ and $B$ are totally
disjoint, i.e., that there is no edge directly linking a vertex of
$A$ to a vertex of $B$. We are interested in the correlations
between these two subsystems. We could consider the correlations
between the intertwiners and edges in $A$ and the ones in $B$.
However, there is no reason that they be correlated. For instance,
they are totally uncorrelated on a pure spin network state. It is
actually more interesting to look at the correlations between $A$
and $B$ induced by the spin network state outside $A$ and $B$.
Intuitively, we would be looking at the correlations between $A$
and $B$ due to the space metric (outside $A$ and $B$). Such a
notion of correlation should be related to the concept of
proximity/distance between $A$ and $B$ in the geometry defined by
the spin network state.

We call $C$ the region exterior to $A$ and $B$. Its boundary is
the union of the boundaries of $A$ and $B$, $\pp C=\pp A \cup \pp
B$ (with opposite orientation). We define the mixed state of the
regions $A,B$ induced by the spin network state in the $C$ region
through a coarse-graining of $C$:
$$
\rho(g,g')\,\equiv\,\int d\mu(g^{(0)}_{e\in C})\,\Theta\psi(\{g_{e\in
\Gamma[C]}\})\,
\overline{\Theta\psi}(\{g'_{e\in \Gamma[C]}\}),
$$
where we recall that $\Gamma[C]=A\cup B \cup \pp A \cup \pp B$.
The coarse-graining map $\Theta$ was defined in the
definition~\ref{def1}. The integration over $g^{(0)}_{e\in C}$
averages over the coarse-graining data. We integrate using the
Haar measure, but we could integrate using a different measure on
the space of holonomies (on the edges of $C$) up to gauge
transformations. If we were to select a single orbit, i.e. a
single choice of coarse-graining data $\{g^{(0)}_e,e\in C\}$, we
would work with the pure coarse-grained state $|\Psi[g^{(0)}_{e\in
C}]\,\ra$.

Explicitly, the state reads:
$$
\rho_{AB}\,\equiv\,
\int d\mu(g^{(0)}_{e\in C})\,
\left|
\int dg \,\psi_\Gamma(g^{(0)}_{e\in C},gg_{e\in\pp C},g_{e\in A},g_{e\in B})
\right|^2\,
|g_{e\in A\cup\pp A}\,g_{e\in B\cup\pp B}\ra\la g_{e\in A\cup\pp
A}\,g_{e\in B\cup\pp B}|,
$$
where the integration over $g$ ensures gauge invariance on the
closed boundary $\pp C$. This state is generally a mixed state
which correlates/entangles the regions $A$ and $B$. Indeed, a
generic state $\psi_\Gamma$ would very unlikely lead to a simple
tensor product state $\rho_{AB}$.

The important point is that from the point of view of $C$ and $\pp
C$, the two regions $A$ and $B$ are behind the same boundary $\pp
C$ and nothing says that they are indeed separated. If we were
given a particular state $\psi_\Gamma$, we could now compute the
correlation/entanglement between $A$ and $B$.  Then we could
define a notion of distance such that the two regions were close
or far away from each other depending whether the correlation in
$\rho_{AB}$ was strong or weak.

\medskip

Let us see how this works on a pure spin network state
$|\psi_\Gamma\ra=|\Gamma,\v{\jmath},\vio\ra$. The state of the
region $C$ is defined by the intertwiners outside $A$ and $B$,
$\bigotimes_{v\in C} \ii_v$. We coarse-grain the whole region $C$
to obtain the boundary state on $\pp C= \pp A \cup\pp B$. For a
given set of holonomies $\{g^{(0)}_e,e\in C\}$ on the edges of
$C$, the resulting intertwiner on $\pp C$ is:
\be
\ii[g^{(0)}_{e\in C}]\,\equiv\,
\int_{\SU(2)}dg\, \tr_{e\in C}\,\left[
\bigotimes_{v\in C} \ii_v \,
\bigotimes_{e\in C} D^{j_e}(g^{(0)}_e) \,
\bigotimes_{e\in \pp C} D^{j_e}(g)\right].
\ee
The coarse-grained state is then the mixed state given by the
density matrix:
\be
\rho_{AB}\,\equiv\,\rho_{\pp C}=
\int \prod_{e\in C} dg^{(0)}_{e\in C}\,
\left|\ii[g^{(0)}_{e\in C}]\right\ra\left\la \ii[g^{(0)}_{e\in C}]\right|,
\ee
where we integrate using the Haar measure. We could integrate
using a different measure on the space of holonomies (on the edges
of $C$) up to gauge transformations. If we were to select a single
orbit, i.e a single choice of coarse-graining data
$\{g^{(0)}_e,e\in C\}$, we would work with the pure state
$|\ii[g^{(0)}_{e\in C}]\,\ra$.

This boundary state lives in the space of intertwiners on $\pp C$,
$\hh_{\pp C}=\Int\left(\otimes_{e\in\pp C} V^{j_e}\right)$. If we
were looking at the boundary states of the two separated regions
$A$ and $B$, we would naturally write a state in $\hh_{\pp A}
\otimes \hh_{\pp B}$. The key point here is that the space of intertwiners on $\pp A \cup \pp B$
is actually
{\it not} the tensor product of the space of intertwiners on $\pp
A$ with the space of intertwiners on $\pp B$, but is strictly
larger:
$$
\hh_{\pp A} \otimes \hh_{\pp B} \subset \hh_{\pp C}.
$$
Intuitively, one could say that the outside metric described by
the state of the region $C$ leads to correlations between the two
regions $A$ and $B$. For instance, it is the metric in the outside
region $C$ which defines the relative position of $A$ and $B$ and
not the metric inside the regions $A$ and $B$ themselves.

\begin{figure}[t]
\psfrag{A}{$A$}
\psfrag{B}{$B$}
\psfrag{C}{$C$}
\psfrag{dA}{$\pp A$}
\psfrag{dB}{$\pp B$}
\psfrag{j}{$j$}
\begin{center}
\includegraphics[width=8cm]{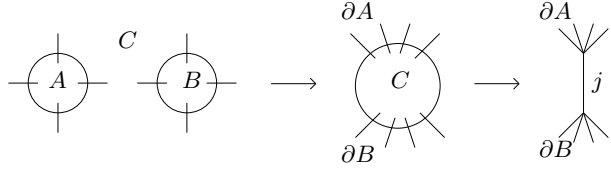}
\end{center}
\caption{\label{jlink}
\small{
Considering two regions $A$ and $B$, and $C$ being the rest of the
spin network, we coarse-grain $C$ to derive the state induced on
the boundary $\pp C=\pp A\cup \pp B$. Finally, separating $\pp A$
and $\pp B$, the resulting intertwiner state can be decomposed in
a basis labelled by a spin $j$ living on an ``internal"/fictitous
link between $A$ and $B$. }}
\end{figure}

Mathematically, $\Int(\otimes_{e\in \pp C} V_{j_e})$ can be
decomposed as the direct sum:
\bes
\label{IntStructure}
\Int(\otimes_{e\in \pp C} V_{j_e}) &=&
\bigoplus_j \Int(V_j\otimes \bigotimes_{e\in \pp A} V_{j_e})
\otimes \Int(V_j\otimes \bigotimes_{e\in \pp B} V_{j_e}) \\
&\equiv& \bigoplus_j \hh^j_{\pp A}\otimes \hh^j_{\pp B} \equiv
\bigoplus_j \hh^j_{AB}. \nonumber
\ees
This decomposition can be easily described pictorially, as shown
in fig.\ref{jlink}, with a fictitious new edge linking $A$ and $B$
and carrying the representation $j$. This corresponds to unfolding
the intertwiner in $\Int(\otimes_{e\in \pp C}V^{j_e})$ on a graph
with two vertices corresponding to $\pp A$ and $\pp B$ but with an
edge linking these two vertices. This edge/link is required by the
$\SU(2)$ invariance of the original intertwiner.
The Hilbert space $\hh_{\pp A}\otimes\hh_{\pp B}$  only
corresponds to the subspace $\hh^{0}_A\otimes\hh^{0}_B$ with the
internal link labelled by the trivial representation $j=0$. In
general, the boundary state $\ii_{AB}=\ii[g^{(0)}_{e\in C}]$ will
not belong to that particular subspace, but will expand on all
possible $j$'s.

In \cite{ourbh}, we studied the case where the state on $\pp C$
was the totally mixed state, the density matrix $\rho_{\pp C}$
being proportional to the identity $\Id$ on $\hh_{\pp C}$. From
these results, we expect the correlations between $A$ and $B$ to
heavily depend on the dominant value of the internal link $j$.
More precisely, we expect the correlations to increase with $j$.
We will study the precise behavior in the next section.

We would like to stress once more that we are looking at the
correlations and entanglement between $A$ and $B$ as induced by
the outside metric i.e the complementary region $C$ to $A\cup B$
in the graph $\Gamma$. Indeed, depending on the values of its
coefficients, the spin network state $|\psi_\Gamma\ra=\sum_{\vio}
c_\vio \,\left|\ii_{v\in A}\ii_{v\in B}\ii_{v\in C} \right\ra $
may or may not entangled the regions $A$ and $B$ i.e be entangled
across the Hilbert space $\otimes_{v\in A}\Int_v \otimes
\otimes_{v\in B}\Int_v$. This is not the correlation/entanglement
which we are investigating: we are looking at the correlations
between $A$ and $B$ induced by the intertwiners $\otimes_{v\in C}
\ii_v$. We expect these latter correlations to be related to the distance between the regions $A$ and
$B$. This will be discussed in more details in the final section.

This provides a concrete basis for the ideas underlined in
\cite{nonlocal} that two regions of the spin network close to each
other in some given embedding of the spin network graph into
$\R^3$ are not necessarily close in the real physical space.
Reversely, two regions of the spin network far to each other in
some given embedding could be strongly correlated and entangled,
which would indicate that that given embedding does not reflect
the structure of the true physical space.

\medskip

We can similarly consider two regions $A$ and $B$ which are linked
by a single edge $e_0$ labelled by a representation $j_0$. That
edge will be inside the region $A\cup B$. Therefore, the value of
the spin $j_0$ will not enter the computation of the global state
$|\ii_{AB}\,\ra= |\ii[g^{(0)}_{e\in C}]\,\ra$: the spin $j$ of the
internal/fictitious link between $A$ and $B$ is completely
independent of the spin $j_0$. Intuitively, one could say that the
link $j_0$ reflects the correlations between $A$ and $B$ as seen
from the inside geometry of $A\cup B$, while the link $j$
describes the correlations between $A$ and $B$ as induced by the
outside geometry (as seen by an observer outside $A$ and $B$).
These considerations can be generalized to the situation where $A$
and $B$ are directly linked through several edges.

\subsection{Bounds on entanglement}
\label{calculation}

As we have shown above, all correlation and entanglement
calculations within a spin network can be reduced to the following
simple set-up. We consider the two systems $A$ and $B$. The
Hilbert space attached to $A$ is the tensor product of the
representation spaces $V_{j_i},\, i\in A$, technically
corresponding to all edges in the boundary of the region $A$. We
similarly define the Hilbert space attached to the system $B$:
$$
\hh_A=\,\bigotimes_{i\in A}V^{j_i}, \qquad
\hh_B=\,\bigotimes_{i\in B}V^{j_i}.
$$
Our state is an intertwiner in the full tensor product
$\bigotimes_{i\in A,B}V^{j_i}$ i.e a singlet or spin-0 state in
$\hh_A\otimes \hh_B$. More precisely, we can decompose the tensor
products in direct sums of irreducible representations:
$$
\hh_X=\bigotimes_{i\in X}V^{j_i}
\,=\,\bigoplus_k V^k\otimes D^X_k,
$$
for both systems $X=A,B$. The space $D^X_k$ is the degeneracy
space of states with spin $k$. We label the basis vectors of
$D^A_k$ as $|\alpha_k\9$, with $\alpha_k$ running from $1$ to
$d^A_k=\dim D^A_k$. Similarly we introduce a basis $|\beta_k\9$ of
the degeneracy space $D^B_k$.

This decomposition provides us directly with a basis of the
intertwiner space $\Int(\otimes_{i\in A,B}V^{j_i})$. Indeed, we
have
$$
\hh_A\otimes\hh_B\,=\,\bigoplus_{k,l} \left(V^k\otimes V^l\right)\,\otimes
\left(D^A_k\otimes D^B_l\right).
$$
Spin-0 states are necessarily states with the spins $k$ and $l$ of
the systems $A$ and $B$ are equal: $k=l$. Therefore, we get:
\be
\hh^0_{AB}\,\equiv\,
\Int(\otimes_{i\in A,B}V^{j_i})\,=\,
\bigoplus_{j} \left(D^A_j\otimes D^B_j\right).
\ee
A basis of the intertwiner space is thus given by the vectors
$|j,\alpha_j,\beta_j\9$. The representation label $j$ is the same
one as in the previous section and is attached to the
fictitous/internal link between the regions $A$ and $B$. It
carries the correlation and entanglement between $A$ and $B$.
Explicitly, these basis states read:
$$
|j,\alpha_j, \beta_j\ra\,=\,
 |j\9_{V^{j}_A\otimes V^{j}_B}\otimes|\alpha_j\9_{D_{j}^A}\otimes|\beta_j\9_{D_{j}^B}
 =\frac{1}{\sqrt{2j+1}}\sum_{m=-j}^{j}
(-1)^{j-m}|j,-m,\alpha_j\9_A\otimes|j,m,\beta_j\9_B.
$$

It was shown in \cite{ourbh} that for a state of the following
type,
$$
\rho=
\sum_j \sum_{a_j,b_j}
\om^{(j)}_{\alpha_j \beta_j}
|j\9\6 j|\otimes |\alpha_j\9\6
\alpha_j|_{D_j^A}\otimes|\beta_j\9\6 \beta_j|_{D_j^B},
\qquad
\sum_{j,\alpha_j ,\beta_j}\om^{(j)}_{\alpha_j \beta_j}=1,
$$
the entanglement between $A$ and $B$ is:
\be
\Ee_\rho(A|B)=\sum_j \log(2j+1)\sum_{\alpha_j ,\beta_j}\om^{(j)}_{\alpha_j
\beta_j}.
\ee
In words, we compute the average of $\log(2j+1)$ over the
probability distribution of the spin $j$ in the state $\rho$.

In particular, \cite{ourbh} considered the totally mixed state,
$$
\rho_0=
\f{1}{N}\sum_j \sum_{a_j,b_j}
|j\9\6 j|\otimes |\alpha_j\9\6
\alpha_j|_{D_j^A}\otimes|\beta_j\9\6 \beta_j|_{D_j^B},
$$
with $N=\dim\hh^0_{AB}=\sum_j d_j^A d_j^B$. For this special
state, the entanglement takes the simple form:
$$
\Ee_{\rho_0}(A|B)=\f{\sum_j d_j^A d_j^B\log(2j+1)}{\sum_j d_j^A
d_j^B},
$$
that is the average of $\log(2j+1)$ over the intertwiner space
$\hh^0_{AB}$.

\medskip

Let us now analyze more carefully the case where the spin $j$
takes a single fixed value. Then states of the previous type,
$$
\rho=\sum_{a_j,b_j}
\om_{\alpha_j \beta_j}
|\alpha_j\9\6 \alpha_j|_{D_j^A}\otimes|\beta_j\9\6
\beta_j|_{D_j^B},
$$
are states that do not carry any entanglement across the
degeneracy spaces. All the entanglement is contained in the spin
$j$. Indeed applying the previous formula, we find exactly:
$$
\Ee_{\rho}(A|B)=\log(2j+1).
$$
If we look allow for arbitrary pure states possibly entangling the
degeneracy spaces,
$$
|\psi\ra\,=\,
\sum_{\alpha_j \beta_j} c_{\alpha_j \beta_j} \,|j,\alpha_j, \beta_j\ra,
$$
then the entanglement is obviously bounded by:
$$
\Ee_{\psi}(A|B)\,\le\,
\log(2j+1)+\log(d_j),
\qquad
d_j=\min(d_j^A,d_j^B),
$$
where the extra term $d_j$ accounts for the contribution from the
degeneracy spaces.

Finally, we allow for arbitrary pure states with possible
superpositions of different values of the spin $j$,
$$
|\psi\ra\,=\,
\sum_j\sum_{\alpha_j \beta_j} c_{j,\alpha_j \beta_j} \,|j,\alpha_j,
\beta_j\ra.
$$
The entanglement is then bounded by
\be
\Ee_{\psi}(A|B)\leq \log d,
\qquad d=\sum_j (2j+1)d_j.
\ee
It is easy to see that $d\leq \sqrt{N}$, where $N=\dim
\hh^0_{AB}$.

\medskip

The definition of the entanglement ensures that in all cases the
maximal possible entanglement is contained in pure states, so that
the bounds derived above hold for arbitrary statistical mixtures.
All these bounds can actually be saturated by pure states.
Moreover, for higher dimension (of the degeneracy spaces), random
pure states are in average nearly maximally entangled
\cite{MaxEnt}.

%
%

\section{Correlations and Distances: a Relational Reconstruction of the Space Geometry?}
\label{TheIdea}

Our goal is to understand the (quantum) metric defined by a spin
network state, without referring to any assumed embedding of the
spin network in a (background) manifold. We support the basic
proposal that a natural notion of distance between two vertices
(or more generally two regions) of that spin network is provided
by the correlations between the two vertices induced by the
algebraic structure of the spin network state. Two parts of the
spin network would be close if they are strongly correlated and
would get far from each other as the correlations weaken.

Our set-up is as follows. We consider two (small) regions, $A$ and
$B$, of the spin network. The distance between them should be
given by the (quantum) metric outside these two regions. Thus we
define the correlations (and entanglement) between $A$ and $B$
induced by the rest of the spin network. This should be naturally
related to the (geodesic) distance between $A$ and $B$.

A first inspiration is quantum field theory on a fixed background.
Considering a (scalar) field $\phi$ for example, the correlation
$\la \phi(x)\phi(y)\ra$ between two points $x$ and $y$ in the
vacuum state depends (only) on the distance $d(x,y)$ and actually
decreases as $1/d(x,y)^2$ in the flat four-dimensional Minkowski
space-time. Reversing the logic, one could measure the correlation
$\la \phi(x)\phi(y)\ra$ between the value of a certain field
$\phi$ at two different space-time points and define the distance
in term of that correlation.

Indeed just as the correlations in QFT contain all the information
about the theory and describes the dynamics of the matter degrees
of freedom, we expect in a quantum gravity theory that the
correlations contained in a quantum state to fully describe the
geometry of the quantum space-time defined by that state.

\medskip

Another inspiration is the study of spin systems, in condensed
matter physics and quantum information \cite{sad,qpt1,qpt2}. Such
spin systems are very close mathematically and physically to the
spin networks of LQG. The key difference is that spin systems are
physical systems embedded in a fixed given background metric
(usually the flat one) while spin networks are supposed to define
that background metric themselves.

Looking at spin systems, we notice that the (total) correlation
between two spins usually obey a power law with respect to the
distance between these two spins. We would like to inverse this
relation in the context of LQG: using a similar power law, we
could reconstruct the distance between two ``points" in term of a
well-defined correlation between the corresponding parts of the
spin network.

The entanglement is trickier to deal with. The entanglement
between two spins usually decreases with the distance, but it
would (almost) vanish beyond the few nearest neighbors. This is
due to  {\it monogamy of entanglement}
\cite{monogamy}: if two systems are
strongly entangled, then none of of them can be strongly
correlated with any other system\footnote{Moreover, a state may be
entangled, while any
 (or some) pair of its subsystems are only correlated, but not entangled. It
 can be seen easily in the following two examples. Consider
 the state of three qubits (the so-called GHZ state, \cite{per, info})
 \be
 |\rm{GHZ}\9=|000\9+|111\9,
 \ee
where we use the conventional $0,1$ basis and ignored the
normalization factors. This is one of the inequivalent maximally
entangled states of three qubits \cite{vidal}, but it is easy to
check that no two subsystems are entangled (even if they are
maximally correlated in the $0,1$ basis). On the other hand,
\be
|w\9=|0\9|0\9|0\9+|1\9(\alpha|0\9|1\9+\beta|1\9|0\9),
\ee
is unentangled across the subsystems 2 and 3, but the two other
pairs are entangled.}.

Interest in the role of entanglement in quantum phase transitions
\cite{sad}, which are characterized by non-analyticity in various
 quantities at certain values of externally-varied parameters at zero
 temperature, was one of the driving forces behind the investigations of entanglement in many-body systems.
  The most popular models are Hamiltonian lattice spin systems in
 one, two and three spatial dimensions. Recently it was shown in various models
 that quantum phase transitions of both first and second kinds
 are signalled by a non-analyticity of the  entanglement
 measures \cite{qpt1,qpt2}.
While the states considered are most typically the ground or
thermal states of the lattice Hamiltonian operator, some of the
results may be usefully applied to the attempts to reconstruct
geometry from the spin networks. Those states typically are
multipartite entangled states. One of the results that are
relevant to us is that the two-qubit entanglement measures vanish
very rapidly with the distance: they may become zero at the
distance of four or five sites, and sometimes even the
third-nearest neighbors entanglement may be zero. At the same
time, the correlations may be strong, and at the critical point
the correlation length may actually diverge. On the other hand,
the overall state is typically entangled, so it is the
multipartite entanglement \cite{ros} that becomes dominant.

Thus, it is likely that the entanglement between two parts of the
spin networks could only be a good notion of distance when the two
systems are very close. Anyway, we only expect the entanglement on
a spin network to be of the same order as the classical
correlation for a pure spin network state with few vertices. We
expect that as soon as we need to coarse-grain over large number
of vertices and edges or if we work on superpositions of spin
network states ({\it weave states}), then the classical
correlation would be much larger than the bipartite
entanglement\footnote{A suitable bipartite measure of entanglement
that reflects its multipartite nature is
\textit{localizable entanglement} (LE) \cite{le}, which as the
maximal amount of entanglement that may be contained in the
bipartite subsystems, on average, by doing local measurements on
other subsystems.  For example, the LE between any two pairs of
qubits in the GHZ state is maximal. The LE leads to the definition
of the entanglement length, which sets a typical scale of its
decay. Its behavior closely resembles the behavior of two-point
correlation functions.}.
\medskip

An obvious criticism to the proposal of reconstructing the
distance from correlations and entanglement is the existence of
EPR pairs: two systems far from each other can still be strongly
entangled. Let us first remind that typical states in quantum
field theories, including the vacuum state, are entangled
\cite{rmp}. Any two points would be naturally entangled in the
vacuum state with the correlation depending on the distance. An
EPR pair would correspond to the creation an anomalously strong
entanglement between two given (space) points. Nevertheless, is
that we are building a statistical notion of distance, which would
only hold in average and in a semi-classical setting. A small
quantum effect like an EPR pair would be averaged out.

A second comment is that EPR pairs concern correlations between
matter degrees of freedom evolving in an already defined geometry
while we are interested of correlations between the degrees of
freedom of the (quantum) geometry. These correlations are supposed
to fully describe the geometry at the quantum level. The
equivalent of an EPR pair in our context would be that the
geometry at two points apparently far from each other could be
strongly entangled/correlated. This would actually look as if
there was a wormhole connecting these two points. This is very
close to the idea of the creation of ``non-local links" in a spin
network state as proposed by Markopoulou and Smolin \cite{Lee},
which could explain some features of dark matter in LQG. Indeed,
taking a spin network based on a regular 3d lattice, one can
create a ``non-local" link, labelled with some representation $j$,
between two points far from each other on the lattice. There is no
argument based on locality forbidding the creation of such a link
since the spin network is defined without any reference of a
background metric: the regular lattice  is simply a particular
embedding of the spin network into flat space and there exists
other embeddings for which the two chosen points are arbitrary
close to each other. Nevertheless, this ``non-local" link will
create some strong entanglement/correlation (especially as $j$
grows) between the two points despite their distance on the
lattice: there will effectively be close as if related by a
wormhole.



Finally, one can argue that the creation of entanglement is always
local: two systems are entangled if and only if they were close at
some point in time. Therefore, the correlation/entanglement is
actually still related to a notion of distance, although it seems
more likely to reflect the possibility of being close (or having
been close).

\medskip

Finally, our ``close/correlated" point of view have some simple
implications for LQG, which could be (easily) checked through
analytical or numerical computations. A first lesson is that a low
spin on an edge between two vertices does not necessarily mean
that these two vertices are close, since low spins tend to lead to
small correlations. Reversely, having no direct link between two
vertices does not necessarily mean that these vertices are far,
since the coarse-graining of the rest of the spin network will
create a link between these two vertices which may carry strong
correlations.


The basic set-up for the simplest explicit calculations consists
of two vertices $A$ and $B$ linked by a single edge (our
internal/fictitious link). We fix the spins on all the edges
coming out of $A$ and $B$ and we also fix the internal edge label
to a given spin $j$. The state of the vertex $A$ is defined as an
intertwiner state in $\Int(V^j\otimes\bigotimes_{e\in \pp A}
V^{j_e})$ while the vertex $B$ is described by an intertwiner in
$\Int(V^j\otimes\bigotimes_{e\in \pp B} V^{j_e})$. Following
section \ref{calculation}, we can compute the entanglement and
total correlation between $A$ and $B$. They tend to grow as the
spin $j$ increases, but the exact value heavily depends on the
choice of intertwiners. Then to reconstruct the distance between
$A$ and $B$, it might be more relevant to compute the ratio of the
actual correlation with the maximal correlation between $A$ and
$B$ (to normalize and get rid of possible $d_{j_e}$ factors). On
the other hand, we could compute the average volume corresponding
to the intertwiner states in $A$ and $B$. Then, since the spin $j$
is the area of the cross-section between the two boundary surfaces
dual to the vertices $A$ and $B$, we should be able to reconstruct
a measure of distance from a certain ratio of $\vv_A \vv_B$ by the
area $\aa\sim j$. At the end of the day, we could compare this two
notions of distance, hoping a close match. We could start by
checking a first expectation:
assuming that the $j_e$'s are more or less all equal, the vertices
$A$ and $B$ would be close if $j$ is large while they get further
away from each other as $j$ decreases. Moreover, if the spins
$j_e$ are large compared to $j$, we expect the vertices to be far.
This should hold for both notions of distance, the ``information
distance" constructed from the correlation (correlations grow with
$j$) and the ``geometric distance" constructed from the volumes
and areas (the cross-section $j$ should grow as the vertices get
closer).

\medskip

A close relationship between these information distance and
geometric distance would strongly support the idea that quantum
information tools should be useful to understand and probe the
quantum  structure of the space(-time) geometry in (loop) quantum
gravity.



\appendix
\section{Coarse-graining and Partial Tracing
Operations}\label{coarse-b}

In this Appendix we show how the coarse-graining relates to the
partial tracing and fill the missing details that were alluded to
in the different places above.

First let us recall the standard definition of a reduced density
operator \cite{ll,per}. To simplify the notations, we consider two
subsystems $A$ and $B$, with Hilbert spaces $\hh_A$ and $\hh_B$,
respectively. Let $\psi(x)$ be a wave function on $\hh_A$,  $\6
x\ket{a}=\phi_a(x)$ label its basis, and $\6
y\ket{b}=\varphi_b(y)$ label the basis of $\hh_B$. The combined
system  has a Hilbert space $\hh=\hh_A\otimes\hh_B$, with the
direct product basis $\ket{ab}\equiv\ket{a}_A\otimes\ket{b}_B$.
 A generic state
$\Psi(x,y)\in\hh$ is given by a linear combination
$\ket{\Psi}=\sum_{a,b}c_{ab}\ket{a}\ket{b}$. If the matrix
elements of the operator $O$ are
\be
\6ab|O|a'b'\9=\sum_{a,a',b,b'}o_{bb'}\delta_{aa'},\label{mateld}
\ee
 then its expectation value is given by
\be
\6\Psi|O|\Psi\9=\sum_{a,b,b'}o_{bb'}c_{ab}\bar{c}_{ab'}=\tr(o\rho_\Psi^B),\label{expred}
\ee
where the reduced density operator
$\rho_\Psi^B\equiv\tr_B\rho_\Psi$ is obtained by tracing out the
subsystem $A$, $\rho^B_{bb'}=\rho_{ab,ab'}$. In the coordinate
basis the operator $O$ is given by
\be
O(x,y;x',y')=o(y,y')\delta(x-x'),\label{matelc}
\ee
 so thanks to the orthonormality of the functions $\phi_a$ the reduced density
 operator is
\be
 \rho^B_\Psi(y,y')=\int dx
 dx'\Psi(x,y)\overline{\Psi}(x',y')\delta(x-x')=c_{ab}c_{ab'}\varphi_b(y)\bar{\varphi}_{b'}(y'),\label{rhoc}
\ee
and
\be
\6\Psi|O|\Psi\9=\int\! dy dy'o(y,y') \rho^B_\Psi(y,y')\label{contexp}.
\ee

\medskip

In the language of the intertwiners a partial tracing does not
present any new features. Taking an intertwiner
$|\vio\9=\otimes_{v\in\Gamma}|\ii_v\9$ and tracing out $A$
produces a basis state of $B=\breve{A}$, $|\vio_B\9=\otimes_{v\in
B}|\ii_v\9\in\hh^0_B \equiv\otimes_{v\in B}\Int_v$, with a
corresponding expressions for the reduced density matrices of
general states.

 Expectation values of the operators that depend
only on the region $B$ are calculated according to
Eq.~(\ref{expred}). For example, the volume operator \cite{lqg} is
a sum of vertex operators, with each vertex operator being a
function of the self-adjoint right-invariant vector fields. The
isomorphism between the spin network states and zero-angular
momentum states of the corresponding abstract spin system leads to
the  form of its matrix elements that is analogous to
Eq.~(\ref{mateld}) and an expectation value of the form of
Eq.~(\ref{expred}). In particular, if $B$ contains a single vertex
$v=1$ and the operator's  matrix elements are
$(O_1)_{\vio\vjo}=o(\ii_1,\jj_1)\prod_{v=2}^V\delta_{\ii_i\jj_i}$,
then for a generic pure state
$|\Psi\9=\sum_{\vio}s_\vio|\vio\9\in\hh^0_\Gamma$
\be
\6\Psi|O_1|\Psi\9=\sum_{\vio,\jj_1}o(\ii_1,\jj_1)s_{\ii_1\ii_2\ldots\ii_V}\bar{s}_{\jj_1\ii_2\ldots\ii_V}.\label{correct}
\ee

Coarse-grained intertwiners have an obvious relation with
partially traced ones. The total coarse-graining of $A$ is given
by
\be
\ii_A=\tr_{v\in A}\!\!\bigotimes_{e\in\inte(A)}\!
D^{j_e}(\Id)\cdot\bigotimes_{v\in A}\ii_v\in\hh_{\pp A},
\ee
where $\tr_{v\in A}$ denotes the summation over pairs of indices
pertaining to the vertices in the region $A$. It turns a
normalized basis state
$|\vio_A\9\otimes|\vio_B\9\in\hh^0_\Gamma=\otimes_{v\in\Gamma}\Int_v$
into a non-normalized state (see Sec.~\ref{probcoarse} for a
discussion) $|\ii_A\9\otimes|\vio_B\9\in\hh^0_{\Gamma[A]}=\hh_{\pp
A}\otimes\hh^0_B$. It can be decomposed as
\be
|\ii_A\9=\sum_\alpha c^{\vio_A}_\alpha|\ii_A^\alpha\9,
\ee
where the intertwiners $\ii_A^\alpha$,
$\alpha=1,\ldots,\dim\hh_{\pp A}$, form the orthonormal basis of
$\hh_{\pp A}$. A general coarse-graining procedure of
Eq.~(\ref{def2}) leads to the same equation, but with the
coefficients depending on the holonomies $\{g_e\}_{e\in\inte(A)}$.
A  normalized pure state
$|\Psi\9=\sum_{\vio}s_\vio|\vio\9\in\hh^0_\Gamma$ becomes a
normalized  pure state
\be
|\Psi_{[A]}\9=\sum_{\vio_A,\vio_B,\alpha}s_{\vio_A\vio_B}c^{\vio_A}_\alpha|\ii_A^\alpha\9|\vio_B\9/\sqrt{\|\ii_A\|}
=\sum_{\vio}s_\vio|\tilde{\ii}_A\9|\vio_B\9\in\hh_{\pp
A}\otimes\hh_B^0.
\ee

Cylindrical functions do not allow a natural separation of
variables in $O_1(g,g')$ that is analogous to Eq.~(\ref{matelc}),
since there is no relation between the number of the intertwiners
$\ii_v$, $v=1,\ldots,|V|$, that define the tensor product
structure of $\hh^0_\Gamma$, and the number of edges, that define
the structure of $H_\Gamma=L^2(\SU(2)^{E}/\SU(2)^{V})$. To obtain
a wave functional that corresponds to $|\vio_{\breve{A}}\9$ one
needs to contract with the representations $D^{j_e}$ that
corresponds to all the edges of $\Gamma[A]$. Hence the
coarse-graining procedures that were described above allow to
introduce the reduced subsystems in the language of cylindrical
functions. From the Definition 1 a total coarse-graining of a spin
network state $\phi_\vio(g)=\6g|\Gamma,\vec{\jmath},\vio\ra$ over
a region $A$ results in $\phi_{\vio [A]}(g)\equiv\6
g|\tilde{\ii}_A\vio_B\9$, where
$|\vio_B\9=\otimes_{v\in_{\breve{A}}}|\ii_v\9\in\hh^0_B$. It is
given explicitly by
\be
\phi_{\vio [A]}(g)=\tr\!\!\! \bigotimes_{e\notin \inte(A)}\!\!D^{j_e}(g_e)\cdot
\mm,\qquad \mm=\tr_{v\in A}\!\!\! \bigotimes_{e\in\inte(A)}\!\!\!
D^{j_e}(\Id)\cdot\bigotimes_{v\in\Gamma}\ii_v/\sqrt{\|\ii_A\|}=\tilde{\ii}_A\otimes\vio_B.
\ee
Consider for simplicity a pure state $|\Psi\9\in\hh^0_\Gamma$.
Mixed states are treated analogously. The following lemma
establishes that the coarse-grained states $\phi_{\vio [A]}(g)$
play a role of the basis wave functions $\varphi_b(g)$ in the
standard partial tracing.
\begin{lemma}
A ``coordinate" expression of the  reduced density matrix
$(\rho_\Psi^B)_{\vio_B\vjo_B}=\sum_{\vio_A}s_{\vio_A\vio_B}\bar{s}_{\vio_A\vjo_B}$
is given by
\be
\rho_\Psi^B(g,g')=
\sum_{\vio,\
\vjo}s_{\vio_A\vio_B}\overline{s}_{\vio_A\vjo_B}\!\int
(dg)^{E_A}(dg')^{E_A}\!\prod_{e\in
\inte(A)}\!\delta(g_e,g'_e)\phi_{\vio
}(g)\overline{\phi}_{\vjo }(g')/\|\ii_A\|=\sum_{\vio,\
\vjo}s_{\vio_A\vio_B}\overline{s}_{\vio_A\vjo_B}\phi_{\vio[A]
}(g)\overline{\phi}_{\vjo[A] }(g'), \label{defred}
\ee
where $|\vio\9=|\vio_A\otimes\vio_B\9$, and
$|\vjo\9=|\vio_A\9\otimes|\vjo_B\9$.
\end{lemma}
We give a proof for $B=\breve{A}$ containing a single vertex
$v=1$. What is needed to be shown is that the expectation value of
the vertex operator $O_1$ is indeed given by the formula analogous
to Eq.~(\ref{rhoc}) and that that Eq.~(\ref{defred}) holds.

 The total coarse-graining over $A=\ext(v_1)$ leads to
$\phi_{\vio[A]}(g)$
with
\be
\mm=\tr\!\!\bigotimes_{e\in \inte(A)} D^{j_e}(\Id)
\cdot \bigotimes_{v\in A} \ii_v
\in \hh_{\pp A}.
\ee
The  functional matrix elements of
$(O_1)_{\vio\vjo}=o(\ii_1,\jj_1)\prod_{v=2}^V\delta_{\ii_i\jj_i}$
are
\be O_1(g,g')\equiv
\6g|\vio\9\6\vio~|O_1|\vjo~\9\6\vjo~|g'\9=\sum_{\vio,
\jj_1}\6g|\ii_1\ii_2\ldots \ii_V \9
o(\ii_1,\jj_1)\6\jj_1\ii_2\ldots \ii_V|g'\9,
\ee and
the totally coarse-grained over $A$ matrix elements can be defined
as
\be O_1^B(g,g')=\sum_{\vio, \jj_1}\phi_{\vio[A] }(g)
o(\ii_1,\jj_1)\overline{\phi}_{\vjo[A] }(g'),
\ee
where $|\vjo\9\equiv|\jj_1\ii_2\ldots\ii_V\9$. Hence to get the
expectation value $\6O_1\9_\Psi$ of Eq.~(\ref{correct}) the
reduced density matrix should be defined as
\be
\rho_\Psi^B(g,g')=\sum_{\vio,\
\vjo}s_{\vio_A\vio_B}\overline{s}_{\vio_A\vjo_B}\phi_{\vio[A]
}(g)\overline{\phi}_{\vjo[A] }(g'),\label{goal}
\ee
and this expression equals the first formula on the lhs of
Eq.~(\ref{defred}) due to Eq.~(\ref{matnorm}).  

The states $|\vio\,\9$ are the eigenstates of the area operators
\cite{lqg} with the eigenvalues that depend only on the
representation labels $j_e$ of the edges that are cut by the
surface. Hence if the region $A$ is defined as an interior  of
some closed surface $S$, the above definition of the reduced
operators is consistent with the expectation values of this
operator that are taken on a full state.

Another family of operators that may be of interest is given by
their functional representation as
\be F(g,g')=\prod_{e\in
E_A}D^{j_e}(g_e)D^{j_e}(g'_e)\cdot M\prod_{\check{e}\in
\ext(A)}\delta(g_{\check{e}},g'_{\check{e}})\equiv
F^B(g,g')\prod_{\check{e} \in
\ext(A)}\delta(g_{\check{e}},g'_{\check{e}}).\label{oper2}
 \ee


\end{document}